%% file: JSS2652_Manuscript.tex
\documentclass[article, shortnames]{jss}

\DeclareGraphicsExtensions{.png,.pdf,.jpg}
\graphicspath{{./Figures/}{./}}
\usepackage{verbatim}
\usepackage{bm, amsmath, amssymb}
\usepackage[toc,page]{appendix}
\usepackage[none]{hyphenat}
\usepackage{caption}
\usepackage{subcaption}
\usepackage{multirow}

\emergencystretch=\maxdimen
\author{Jingjing Yang \\University of Michigan \And 
        Peng Ren \\Suntrust Banks, Inc.}
\title{\pkg{BFDA}: A \proglang{MATLAB} Toolbox for Bayesian Functional Data Analysis}

\Plainauthor{Jingjing Yang, Peng Ren} 
\Plaintitle{A MATLAB Toolbox for Bayesian Functional Data Analysis } 
\Shorttitle{\pkg{BFDA}: Bayesian Functional Data Analysis} 

\newcommand{\bXi}{{\bm X}_{\bm{t}_i}}

\newcommand{\bZi}{{\bm Z}_{\bm{t}_i}}

\Abstract{
{\sloppy
We provide a \proglang{MATLAB} toolbox, \pkg{BFDA}, that implements a Bayesian hierarchical model to smooth multiple functional data with the assumptions of the same underlying Gaussian process distribution, a Gaussian process prior for the mean function, and an Inverse-Wishart process prior for the covariance function. This model-based approach can borrow strength from all functional data to increase the smoothing accuracy, as well as estimate the mean-covariance functions simultaneously. An option of approximating the Bayesian inference process using cubic B-spline basis functions is integrated in \pkg{BFDA}, which allows for efficiently dealing with high-dimensional functional data. Examples of using \pkg{BFDA} in various scenarios and conducting follow-up functional regression are provided. The advantages of \pkg{BFDA} include: (1) Simultaneously smooths multiple functional data and estimates the mean-covariance functions in a nonparametric way; (2) flexibly deals with sparse and high-dimensional functional data with stationary and nonstationary covariance functions, and without the requirement of common observation grids; (3) provides accurately smoothed functional data for follow-up analysis.}
}
\Plainkeywords{functional data analysis, Bayesian hierarchical model, Gaussian process, Inverse-Wishart process, MATLAB } 

\Address{
  Jingjing Yang\\
  Department of Biostatistics\\
  School of Public Health\\
  University of Michigan\\
  1415 Washington Heights\\ 
  Ann Arbor, MI, 48109, U.S.A. \\
  E-mail: \email{yjingj@umich.edu; yjingj@gmail.com}
  }

\begin{document}

\input{Introduction.tex}

\input{Methods.tex}

\input{ExampleScript.tex}

\input{Regression.tex}

\input{Discussion.tex}

\bibliography{BFDA_JSS}

\input{Appendix}
\end{document}

%% file: Introduction.tex
\section{Introduction}
\label{intro}

Since \citet{ramsay1991some} first coined the term ``functional data analysis'' (FDA) for analyzing data that are realizations of a continuous function, many statistical methods and tools have been proposed for FDA. For examples, \citet{graves2010functional} provided both \proglang{R} package \pkg{fda} \citep{fda_rpack}  and \proglang{MATLAB} package \pkg{fdaM} \citep{fda_M}
for standard functional data analysis \citep{ramsay2002applied, ramsay2005functional};
\citet{fda_usc} provided \proglang{R} package \pkg{fda.usc} for implementing nonparametric functional data analysis methods \citep{vieu2006nonparametric} with \pkg{fda} \citep{fda_rpack};
 \citet{Yao2005, yao2005functional} developed the key technique of Functional Principal Component Analysis (FPCA) for analyzing sparse functional data, accompanied by the \proglang{MATLAB} package \pkg{PACE} \citep{PACE}; \citet{crainiceanu2010bayesian} proposed insights about implementing the standard Bayesian FDA using \pkg{WinBUGS} \citep{Winbugs}; and \citet{GPFDA_rpack} derived \proglang{R} package \pkg{GPFDA} for applying the Bayesian nonparametric Gaussian process (GP) regression models \citep{shi2011}. 
 However, the smoothing step that constructs functions from noisy discrete data has been neglected by most of the existing FDA methods and tools, where functional representations are integrated in the analyzing models. On the other hand, most of the existing smoothing methods process functional samples individually (e.g., cubic smoothing spline (CSS) and kernel smoothing \citep{green1993nonparametric, ramsay2005functional}), which is likely to induce bias when functional samples are of the same distribution. 

 Here, we provide a \proglang{MATLAB} toolbox \pkg{BFDA} for simultaneously smoothing multiple functional observations from the same distribution and estimating the underlying mean-covariance functions, using a nonparametric Bayesian Hierarchical Model (BHM) with Gaussian-Wishart processes \citep{yang2016}. This model-based approach borrows strength through modeling the shared mean-covariance functions, thus producing more accurate smoothing results than the individually smoothing methods \citep{yang2016}. Moreover, \pkg{BFDA} is flexible for analyzing sparse and dense functional data without the requirement of common observation grids, suitable for analyzing functional data with both stationary and nonstationary covariance functions, and efficient for high-dimensional functional data using a Bayesian framework with Approximations by Basis Functions (BABF) \citep{yang_arXiv}. In addition, \pkg{BFDA} provides options for implementing the standard Bayesian GP regression method, conducting Bayesian principal component analysis, and using the \pkg{fdaM} package for follow-up FDA.

In the following context, we first review the BHM and BABF methods in Section \ref{method}, and then provide examples using \pkg{BFDA} with simulated data in Section \ref{exam}. In Section \ref{reg}, we compare the functional linear regression results by \pkg{fdaM} using the smoothed data by \pkg{BFDA} and CSS. Last, we conclude with a discussion in Section \ref{dis}. Details of input options and outputs, as well as example \proglang{MATLAB} scripts of generating the simulation results in this paper, are provided in the Appendix.

%% file: Methods.tex
\section{Methods overview}
\label{method}

\subsection{BHM}
\label{bhm}

Consider functional data $\{X_i(t);\, t\in \mathcal{T}, \, i = 1, 2, \cdots, n\}$ that are generated from the same stochastic process with independent measurement errors. In order to simultaneously smooth all noisy observations and estimate mean-covariance functions, \citet{yang2016} proposed the following Bayesian Hierarchical Model (BHM) with Gaussian-Wishart processes:
\begin{equation}
\label{bhm_mod}
X_i(t) = Z_i(t) + \epsilon_i(t); \; Z_i(\cdot) \sim GP(\mu_Z(\cdot), \Sigma_Z(\cdot, \cdot)),\; \epsilon_i(\cdot) \sim N(0, \sigma^2_{\epsilon});
\end{equation}
$$\mu_Z(\cdot)|\Sigma_Z(\cdot, \cdot) \sim GP\left(\mu_0(\cdot), \frac{1}{c}\Sigma_Z(\cdot, \cdot)\right), \; \Sigma_Z(\cdot, \cdot) \sim IWP(\delta, \sigma^2_sA(\cdot, \cdot)), \; \sigma^2_{\epsilon} \sim IG(a_{\epsilon}, b_{\epsilon}); $$
$$\sigma^2_s \sim IG(a_s, b_s); \; $$
where $\{Z_i(t); i = 1, \cdots, n\}$ denotes the underlying true functional data following the same GP distribution with mean function $\mu_Z(\cdot)$ and covariance function $\Sigma_Z(\cdot, \cdot)$, $IWP$ denotes the Inverse-Wishart process (IWP) prior \citep{dawid1981some} for the covariance function, $IG$ denotes the Inverse-Gamma prior, and $(\mu_0(\cdot), c, \delta, A(\cdot, \cdot), a_{\epsilon}, b_{\epsilon}, a_s, b_s)$ are hyper-parameters to be determined. The IWP prior on $\Sigma_Z(\cdot, \cdot)$ models the covariance function nonparametrically and therefore makes the BHM method suitable for analyzing functional data with unknown stationary and nonstationary covariance structures. The hyper parameter $\sigma_s^2$ provides the flexibility of estimating the scale of the covariance structure in the IWP prior from the data.  

For the hyper-parameter setup, we take $\mu_0(\cdot)$ as the smoothed empirical mean estimate, $c$ as $1$ for the same data variation on the mean function, $\delta$ as $5$ for noninformative prior on the variance function, and determine $(a_{\epsilon}, b_{\epsilon}, a_s, b_s)$ by matching the hyper-prior moments with the empirical estimates. In addition, $A(\cdot, \cdot)$ can be taken as the Mat\'{e}rn correlation kernel for analyzing functional data with stationary covariance (default in \pkg{BFDA}), 
$$
\mbox{Matern}_{cor}(d; \rho, \nu) =  \frac{1}{\Gamma(\nu) 2^{\nu - 1}}
\left(\sqrt{2\nu}\frac{d}{\rho}  \right)^{\nu}  K_{\nu}\left( \sqrt{2\nu}\frac{d}{\rho}\right), \quad d\ge 0,\; \rho>0, \; \nu>0,
$$
where $d$ denotes the distance between two grid points,
$\rho$ is the scale parameter, $\nu$ is the order of smoothness, and $K_\nu(\cdot)$ is the modified Bessel function of the second kind; while as an smoothed empirical covariance estimate for analyzing functional data with nonstationary covariance.

Although the BHM is constructed with infinite-dimensional Gaussian-Wishart processes, practical posterior inference will be conducted in a finite manner, e.g., on the observation grids $\{\bm{t_i}\}$, pooled grid $\bm{t} = \cup_{i=1}^n \bm{t}_i$, or other evaluation grids. For accommodating uncommon observation grids, especially sparsely observed data, BHM evaluates data functions and mean-covariance functions on the pooled grid, while estimating the unobserved functional data by conditioning on the observations (similarly as \pkg{PACE}).

Denoting $X_i(\bm{t}_i)$ by $\bXi$, $Z_i(\bm{t}_i)$ by $\bZi$, $\mu_0(\bm{t})$ by $\bm{\mu}_0$,  $\mu_Z(\bm{t})$ by $\bm{\mu}_Z$, $\Sigma_Z(\bm{t}, \bm{t})$ by $\bm{\Sigma}_Z$, and $A(\bm{t}, \bm{t})$ by $\bm{A}$, BHM conducts Bayesian inference for 
$(\{Z_i(\bm{t})\}, \bm{\mu}_Z, \bm{\Sigma}_Z, \sigma^2_{\epsilon}, \sigma_s^2 )$ by the Monte Carlo Markov Chain (MCMC) algorithm (essentially a Gibbs sampler \citep{geman1984stochastic}) as follows (refer to \citet{yang2016} for details):
\vspace{-0.1 in}
\begin{itemize} 
\item[] Step 0: Set initial values. Set hyper-parameters $(c, \bm{\mu_0}, \nu, \rho, a_{\epsilon}, b_{\epsilon}, a_s, b_s)$. Take
$(\bm{\mu}, \sigma_{\epsilon}^2)$ as the empirical estimates, $\{\bZi\}$ as the raw data $\{\bXi\}$, and $\bm{\Sigma}_Z$ as an identity matrix. 
\item[] Step 1: Conditioning on $\{\bXi\}$ and $(\bm{\mu}_Z,\bm{\Sigma}_Z)$, update $\{Z_i(\bm{t})\}$ from the corresponding conditional multivariate normal (MN) distributions; 
\item[] Step 2: Conditioning on $\{\bXi\}$ and $\{\bZi\}$, 
update the noise variance $\sigma^2_{\epsilon}$ from the conditional Inverse-Gamma (IG) distribution;
\item[] Step 3: Conditional on $\{Z_i(\bm{t})\}$ and $\bm{\Sigma}_Z$, update $\bm{\mu}_Z$ from 
 the conditional MN distribution;
\item[] Step 4: Conditioning on $\{Z_i(\bm{t})\}$ and $\bm{\mu}_Z$, update $\bm{\Sigma}_Z$ from
 the conditional Inverse-Wishart (IW) distribution;
\item[] Step 5:  Conditioning on $\mathbf{\Sigma}_Z$, Update $\sigma_s^2$ from the conditional Gamma distribution.
\end{itemize}
\vspace{-0.1 in}
Specifically, the averages of posterior samples of $\{Z_i(\bm{t}), \bm{\mu}_Z, \bm{\Sigma}_Z\}$ are taken as estimates for functional signals and mean-covariance functions.

In addition, \pkg{BFDA} uses existing \proglang{MATLAB} package \pkg{mcmcdiag} \citep{mcmcdiag} to diagnosis the MCMC convergence by potential scale reduction factor (PSRF) \citep{gelman1992inference}, and implements the method proposed by \citet{yuan2012} with the pivotal discrepancy measures (PDM) of standardized residuals for the goodness-of-fit diagnosis of the assumed model.

\subsection{BABF}
\label{babf}

Because BHM \citep{yang2016} has computational complexity $O(np^3m)$ with $n$ samples, $p$ pooled-grid points, and $m$ MCMC iterations, the method encounters computational bottleneck for analyzing functional data with large pooled-grid dimension $p$. 
To resolve this computational bottleneck issue with high-dimensional data, \pkg{BFDA} enables the option of using the BABF method proposed by \citet{yang_arXiv}, in which the posterior inference in BHM is conducted with approximations using basis functions. Here, we briefly outline the BABF method.  

First select a working grid based on data density, $\bm{\tau} = (\tau_1, \tau_2, \cdots, \tau_L)^\top \subset \mathcal{T}$, $L<<p$, then approximate $\{Z_i(\bm{\tau})\}$ by a system of basis functions (e.g., cubic B-splines). Let $\bm{B}(\cdot) = [b_1(\cdot), b_2(\cdot), \cdots, b_K(\cdot)]$ denote $K$ selected basis functions with coefficients $\bm{\zeta_i} = (\zeta_{i1}, \zeta_{i2}, \cdots, \zeta_{iK})^\top$, then
$Z_i(\bm{\tau}) = \sum_{k = 1}^K \zeta_{ik} b_k(\bm{\tau}) = \bm{B(\tau)} \bm{\zeta_i}$. 
Generally,  $K$ can be selected as $L$, then $\bm{\zeta_i} = \bm{B(\tau)}^{-1} Z_i(\bm{\tau})$, a linear transformation of $Z_i(\bm{\tau})$. Note that even if $\bm{B(\tau)}$ is singular or non-square, $\bm{\zeta_i}$ can still be written as a linear transformation of $Z_i(\bm{\tau})$, e.g., using the generalized inverse \citep{james1978generalised} of $\bm{B(\tau)}$. 

Because $\bm{\zeta_i}$ is a linear transformation of $Z_i(\bm{\tau})$ that follows a MN distribution under the assumptions in Equation~\ref{bhm_mod}, the induced Bayesian hierarchical model for $\{\bm{\zeta_i}\}$ is derived as
\begin{equation}
\label{zeta}
\bm{\zeta_i} \sim MN(\bm{\mu_{\zeta}},\; \bm{\Sigma_{\zeta}}); \; \bm{\mu_{\zeta}} = \bm{B(\tau)}^{-1} \mu_Z(\bm{\tau}); \; \bm{\Sigma_{\zeta}} = \bm{B(\tau)}^{-1} \Sigma_Z(\bm{\tau}, \bm{\tau}) \bm{B(\tau)}^{-\top}.
\end{equation} 
Further, from the assumed priors of $(\mu_Z(\cdot), \Sigma_Z(\cdot, \cdot))$ in Equation~\ref{bhm_mod}, with $\Psi(\bm{\tau}, \bm{\tau})= \sigma_s^2 A(\bm{\tau}, \bm{\tau})$,
the following priors of $(\bm{\mu_{\zeta}}, \bm{\Sigma_{\zeta}})$ are also induced: 
\begin{eqnarray}
\bm{\mu_{\zeta}}| \bm{\Sigma_{\zeta}}  &\sim& MN\left(\bm{B(\tau)}^{-1} \mu_0(\bm{\tau}),\; c\bm{\Sigma_{\zeta}} \right); \label{mu_zeta_prior}\\
\bm{\Sigma_{\zeta}} &\sim& IW(\delta, \; \bm{B(\tau)}^{-1} \Psi(\bm{\tau}, \bm{\tau}) \bm{B(\tau)}^{-\top}). \nonumber  
\end{eqnarray}

Then, the BABF approach has computation complexity $O(nK^3m)$ with the following MCMC algorithm (refer to \citet{yang_arXiv} for details):
\vspace{-0.1 in}
\begin{itemize}
\item[]  Step 0: Set initial values similarly as in BHM. Set hyper-parameters $(c, \bm{\mu_0}, \nu, \rho, a_{\epsilon}, b_{\epsilon}, a_s, b_s)$. Take $(\mu_Z(\bm{\tau}), \Sigma_Z(\bm{\tau}, \bm{\tau}), \sigma^2_{\epsilon})$ as empirical estimates, inducing the initial values for $(\bm{\mu_{\zeta}}, \bm{\Sigma_{\zeta}})$ by Equation~\ref{zeta}.

\item[]  Step 1: Conditioning on $\{\bXi\}$ and $(\bm{\mu_{\zeta}}, \bm{\Sigma_{\zeta}}, \sigma_{\epsilon}^2)$, update $\{\bm{\zeta_i}\}$ from the conditional MN distribution.

\item[]  Step 2: Conditioning on  $\{\bm{\zeta_i}\}$, update $\bm{\mu_{\zeta}}$ and $\bm{\Sigma_{\zeta}}$ respectively from the conditional MN and IW distributions.

\item[]  Step 3: Conditioning on $(\{\bm{\zeta_i}\}, \bm{\mu_{\zeta}}, \bm{\Sigma_{\zeta}})$, approximate $\{\bZi$, $\mu_Z(\bm{t_i})$, $\Sigma_Z(\bm{t_i}, \bm{t_i})$, $\Sigma_Z(\bm{\tau}, \bm{t_i}), \Sigma_Z(\bm{t_i}, \bm{\tau}), \Sigma_Z(\bm{\tau}, \bm{\tau})\}$ by 
$$\bZi = \bm{B(t_i)} \bm{\zeta_i}, \; \mu_Z(\bm{t_i}) = \bm{B(t_i)\mu_{\zeta}}, \;  \Sigma_Z(\bm{t_i}, \bm{t_i}) =  \bm{B(t_i)}\bm{\Sigma_{\zeta}} \bm{B(t_i)}^\top, \; $$
$$\Sigma_Z(\bm{\tau}, \bm{t_i})^\top = \Sigma_Z(\bm{t_i}, \bm{\tau}) =  \bm{B(t_i)}\bm{\Sigma_{\zeta}} \bm{B(\tau)}^\top, \; 
\Sigma_Z(\bm{\tau}, \bm{\tau}) =  \bm{B(\tau)}\bm{\Sigma_{\zeta}} \bm{B(\tau)}^\top .$$

\item[]  Step 4: Conditioning on $\{\bZi\}$ and $\{\bXi\}$, update $\sigma_{\epsilon}^2$ by
the conditional IG distribution;

\item[]  Step 5: Conditioning on $\Sigma_Z(\bm{\tau}, \bm{\tau})$, update $\sigma_s^2$   by the conditional Gamma distribution.
\end{itemize}

As a result, the posterior estimates of $(\{\bm{\zeta_i}\}, \bm{\mu_{\zeta}}, \bm{\Sigma_{\zeta}})$ are given by the averages of the MCMC samples, which are then used to estimate $\{\bZi$, $\mu_Z(\bm{t_i})$, $\Sigma_Z(\bm{t_i}, \bm{t_i})\}$ by the approximation formulas in Step 3. 

\subsection{Basis-function construction}
\label{basis-func}

\pkg{BFDA} uses the existing \proglang{MATLAB} package \pkg{bspline} \citep{bspline} to construct B-spline basis functions, using the optimal knot sequence for interpolation at the working grid $\tau$. The optimal knot sequence is generated by the \proglang{MATLAB} function \verb|optknt| \citep{gaffney1976optimal, micchelli1976optimal, de1977computational}. \citet{yang_arXiv} instructed selecting $\tau$ to represent data densities ($L$ maybe selected by grid search with test data), such as taking the $\left(\frac{100}{L+1}, \cdots, \frac{L\times100}{L+1}\right)$ percentiles of the pooled grid, or the equally-spaced grid for evenly distributed data.

%% file: ExampleScript.tex
\section{Examples with simulated data}
\label{exam}

In this Section, we provide examples of using \pkg{BFDA} to analyze simulated functional data with stationary and nonstationary covariance functions, common and uncommon (sparse) observation grids, as well as random observation grids. The simulation data used for the example results were generated with \code{n}=30, \code{p}=40, \code{au}=0, \code{bu}=$\pi/2$, \code{s}=$\sqrt{5}$, \code{r}=2, \code{nu}=3.5, \code{rho}=0.5, \code{dense}=0.6, and \code{pgrid} as the equally spaced grid over $(0, \pi/2)$ with length $40$. 

\subsection{Simulate functional data}
\label{simfd}

\pkg{BFDA} provides the convenience of generating simulated functional data from the same GP with mean function $\mu(t) = 3sin(4t)$, stationary covariance function $s^2\mbox{Matern}_{cor}(d; \rho, \nu)$, and noises $\sim N(0, (s/r)^2)$ by

\code{
> GausFD_cgrid = sim_gfd(pgrid, n, s, r, nu, rho, dense, cgrid, stat);
}

where \code{pgrid} denotes the pooled grid, \code{n} denotes the number of functional samples, \code{r} denotes the signal to noise ratio  (i.e., the ratio between the signal and noise standard deviations), \code{rho} denotes the Mat\'ern scale parameter, \code{nu} denotes the Mat\'ern order of smoothness. Here, \code{cgrid} is a boolean indicator that controls the output as either common-grid data on \code{pgrid} (\code{cgrid}=1) or uncommon-grid data with a randomly selected proportion (given by \code{dense}) of the full data on \code{pgrid} (\code{cgrid}=0). In addition, \code{stat}=1 specifies simulating stationary data from $GP(3sin(4t), s^2\mbox{Matern}_{cor}(d; \rho, \nu))$, while  \code{stat}=0, specifies simulating data from a nonlinearly transformed GP with mean function $\mu(t) = 3(t+0.5)sin(4t^{2/3})$ and nonstationary covariance function 
$\Sigma(t, t') = s^2(t+0.5)(t'+0.5)\mbox{Matern}_{cor}(|t^{2/3} - t'^{2/3}|;\rho, \nu).$

Let $p$ denote the length of \code{pgrid}.
The output \code{GausFD_cgrid} is a data structure consisted with a cell  of true data ($\text{Xtrue\_cell}_{1\times n}$), a cell of noisy data ($\text{Xraw\_cell}_{1\times n}$), a cell  of observation grids ($\text{Tcell}_{1\times n}$), a true covariance matrix on \code{pgrid} ($\text{Cov\_true}_{p\times p}$), and a true mean vector on \code{pgrid} ($\text{Mean\_true}_{1\times p}$).

\subsection{Analyze stationary data by BHM}
\label{statfd}

\subsubsection{Common grids}
First, we need to setup the required parameter structure by function \code{setOptions_bfda}. For example, to analyze functional data with common observation grids and stationary covariance function by BHM, the structure \code{param} can be set as 

\code{
> param = setOptions_bfda('smethod', 'bhm', 'cgrid', 1, 'mat', 1, 'M', 10000, 'Burnin', 2000, 'w', 1, 'ws', 1);
}

where each parameter is followed by its value, and unspecified parameters are taken as default values (Appendix A.1.). Specifically, \code{smethod='bhm'} denotes using the BHM method; \code{cgrid=1} denotes the analyzed data are of common-grid; \code{mat=1} denotes taking $A(\cdot, \cdot)$ in Equation \ref{bhm_mod} as the Mat\'ern correlation function; \code{M=10000} denotes the number of MCMC iterations; 
\code{Burnin=2000} denotes the number of MCMC burn-ins; \code{w=1} and \code{ws=1} are used to tune the Gamma priors for $\sigma_{\epsilon}^2$ and $\sigma_s^2$.

With both \code{param} and \code{GausFD_cgrid}, we can then call the main function \code{BFDA()} by 

\code{
> [out_cgrid, param] = BFDA(GausFD_cgrid.Xraw_cell, GausFD_cgrid.Tcell, param);
}

for smoothing and estimating the common-grid functional data by BHM. The output structure \code{out_cgrid} contains smoothed estimates for the signals (\code{out_cgrid.Z}), mean function (\code{out_cgrid.mu}), covariance function (\code{out_cgrid.Sigma}), and other parameters in Equation \ref{bhm_mod}, along with the corresponding 95\% point-wise credible intervals (Appendix A.1.). The output argument \code{param} is the updated parameter structure. 

\subsubsection{Uncommon grids}
To apply BHM on stationary functional data of uncommon-grid, e.g., \code{GausFD_ucgrid} generated by 

\code{
> GausFD_ucgrid = sim_gfd(pgrid, n, s, r, nu, rho, dense, 0, stat);	
}

the main function \code{BFDA} can be called by 

\code{
> param_uc = setOptions_bfda('smethod', 'bhm', 'cgrid', 0, 'mat', 1, 'M', 10000, 'Burnin', 2000, 'pace', 1, 'ws', 0.1);
}

\code{
> [out_ucgrid, param_uc] = BFDA(GausFD_ucgrid.Xraw_cell, GausFD_ucgrid.Tcell, param_uc);
}

where \code{cgrid} is set as $0$ in \code{param_uc}.

\subsubsection{Example results}
\begin{figure}[hb]
\begin{center}
\includegraphics[width=0.65\textwidth]{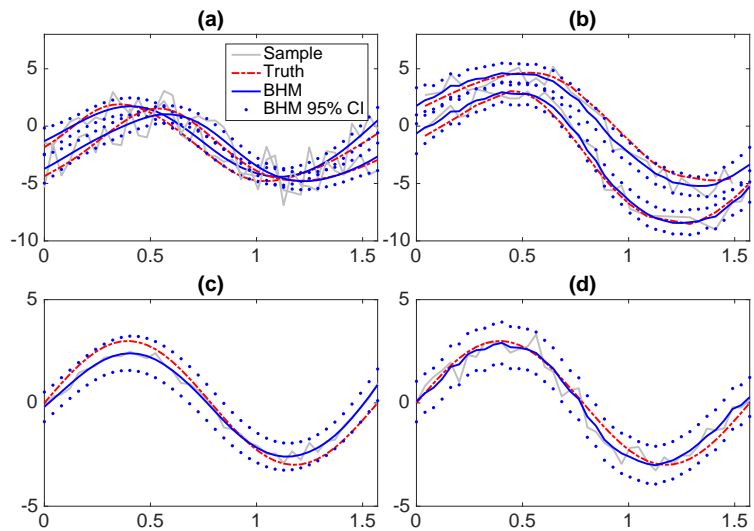}
\caption{Results of analyzing \textbf{Stationary} functional data by BHM: (a) two sample signal estimates with common grids; (b) two sample signal estimates with uncommon grids; (c) mean estimate with common grids; (d) mean estimate with uncommon grids; along with 95\% pointwise credible intervals (blue dots).}
\label{fig:1}
\end{center}
\end{figure}

\begin{figure}[htb]
\begin{center}
\includegraphics[width=0.7\textwidth]{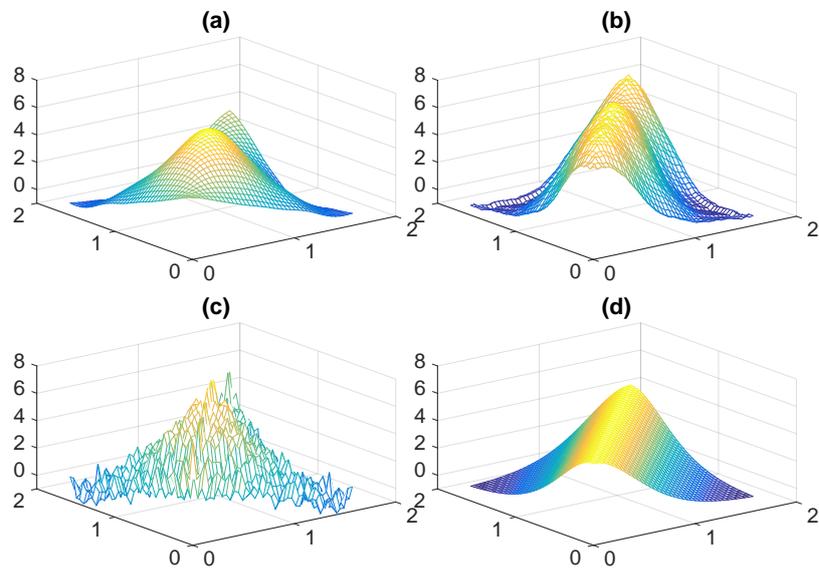}
\caption{Covariance estimates for \textbf{Stationary} functional data: (a) BHM estimate with common grids; (b) BHM estimate with uncommon grids; (c) sample estimate with raw common-grid data; (d) true covariance surface.}
\label{fig:2}
\end{center}
\end{figure}

In Figure \ref{fig:1}(a, b), we show that the smoothed signals by BHM (blue solid) are close to the truth (red dashed), and the coverage probabilities of the 95\% point-wise credible intervals (blue dotted) are $>0.95$, for both scenarios with common and uncommon grids. In addition, the nonparametric mean estimates by BHM (blue solid curves in Figure \ref{fig:1}(c, d)) are also smooth and close to the truth (red dashed), while the corresponding 95\% point-wise credible intervals (blue dotted) have coverage probabilities $>0.9$. In addition, we show that the Bayesian nonparametric covariance estimates in Figure \ref{fig:2}(a, b) are clearly smoother than the sample covariance estimate by using the raw common-grid data in Figure \ref{fig:2}(c), and close to the true stationary covariance in Figure \ref{fig:2}(d). Importantly, although 40\% information is lost for the uncommon-grid scenario, BHM still produces good smoothing and estimation results.  

\clearpage

\subsection{Analyze nonstationary data by BHM}
\label{nstat_fd}

\subsubsection{Common grids}
To apply BHM on functional data with nonstationary covariance function and common grids, e.g., \code{GausFD_cgrid_ns} generated by

\code{
> GausFD_cgrid_ns = sim_gfd(pgrid, n, s, r, nu, rho, dense, cgrid, 0);
}

the main function \code{BFDA()} can be called by  

\code{
> param_ns = setOptions_bfda('smethod', 'bhm', 'cgrid', 1, 'mat', 0, 'M', 10000, 'Burnin', 2000, 'pace', 1, 'ws', 0.01);
}

\code{
> [out_cgrid_ns, param_ns] = BFDA(GausFD_cgrid_ns.Xraw_cell, GausFD_cgrid_ns.Tcell, param_ns);
}

Here, $A(\cdot, \cdot)$ in (\ref{bhm}) is set as the empirical smooth covariance estimate (\code{mat}=0) that is given by \pkg{PACE} \citep{Yao2005, PACE} with \code{pace}=1 (default), or by the sample covariance estimate using CSS smoothed data with  \code{pace}=0. 

\subsubsection{Uncommon grids}
If the nonstationary functional data are collected on uncommon (sparse) grids, e.g., \code{GausFD_ucgrid_ns} generated by

\code{
> GausFD_ucgrid_ns = sim_gfd(pgrid, n, s, r, nu, rho, dense, 0, 0);
}

we only need to set \code{cgrid}=0, \code{mat}=0 in the parameter structure for the common-grid scenario and then call the main function \code{BFDA()} by

\code{
> param_uc_ns = setOptions_bfda('smethod', 'bhm', 'cgrid', 0, 'mat', 0, 'M', 10000, 'Burnin', 2000, 'pace', 1, 'ws', 0.01);
}

\code{
> [out_ucgrid_ns, param_uc_ns ] = BFDA(GausFD_ucgrid_ns.Xraw_cell, GausFD_ucgrid_ns.Tcell, param_uc_ns);
}

where \code{cgrid} is set as $0$ in \code{param_uc_ns}.

\subsubsection{Example results}

\begin{figure}[htb]
\begin{center}
\includegraphics[width=0.65\textwidth]{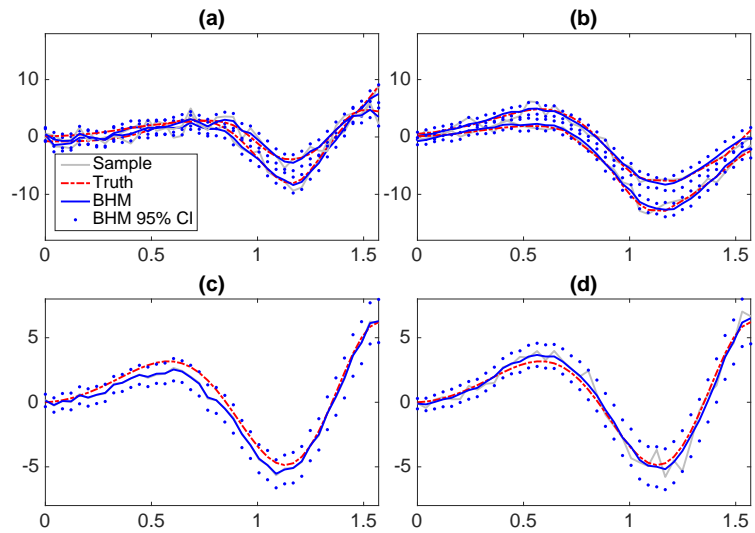}
\caption{Results of analyzing \textbf{Nonstationary} functional data: (a) two sample signal estimates with common grids; (b) two sample signal estimates with uncommon grids; (c) mean estimate with common grids; (d) mean estimate with uncommon grids; along with 95\% pointwise credible intervals (blue dots).}
\label{fig:3}
\end{center}
\end{figure}

\begin{figure}[htb]
\begin{center}
\includegraphics[width=0.7\textwidth]{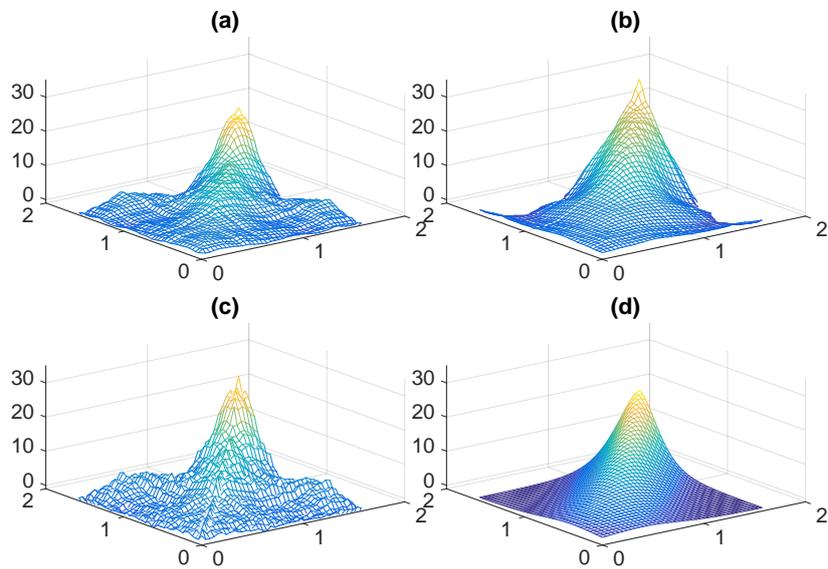}
\caption{Covariance estimates for \textbf{Nonstationary} functional data: (a) BHM estimate with common grids; (b) BHM estimate with uncommon grids; (c) sample estimate with raw common-grid data; (d) true covariance surface.}
\label{fig:4}
\end{center}
\end{figure}

Similarly, as shown in Figures \ref{fig:3} and \ref{fig:4}, the BHM estimates of signals and mean-covariance functions are close to the truth. Specifically, the $95\%$ pointwise credible intervals of the BHM signal estimates have coverage probabilities $>0.95$. Although BHM overestimated the covariance, BHM captured the major covariance structure and produced a smoothed estimate.
The magnitude of the BHM estimate can be tuned by \code{ws}, where a smaller \code{ws} will relatively shrink the magnitude of BHM covariance estimate. We suggest users to tune this parameter according to the magnitude of sample covariance estimate.  

\clearpage

\subsection{Analyze functional data by BABF}

\pkg{BFDA} also provides the convenience to simulate stationary and nonstationary functional data with random observation grids from the same GPs as in Section \ref{simfd}. For example, a structure of functional data \code{GausFD_rgrid}, with \code{n} independent observations and \code{p} random grids per observation (uniformly generated from [\code{au}, \code{bu}], can be generated by

\code{
> GausFD_rgrid = sim_gfd_rgrid(n, p, au, bu, s, r, nu, rho, stat);
} 

where \code{stat=1} specifies simulating from the stationary GP, while \code{stat=0} specifies simulating from the nonstationary GP.

\subsubsection{Stationary data}

To analyze stationary functional data by BABF, simply call the main function \code{BFDA()} by:

\code{
> param_rgrid = setOptions_bfda('smethod', 'babf', 'cgrid', 0, 'mat', 1, 'M', 10000, 'Burnin', 2000, 'm', m, 'eval_grid', pgrid, 'ws', 1);
}

\code{
> [out_rgrid, param_rgrid]= BFDA(GausFD_rgrid.Xraw_cell, GausFD_rgrid.Tcell, param_rgrid);
}

where the working grid $\bm{\tau}$ will be set as the equally spaced quantiles of the pooled grid by default, with length \code{m} (when \code{tau} is not initialized in \code{param_rgrid}).

\subsubsection{Nonstationary data}
For nonstationary functional data, e.g., \code{GausFD_rgrid_ns} generated by

\code{
> GausFD_rgrid_ns = sim_gfd_rgrid(n, p, au, bu, s, r, nu, rho, 0);
}

we can call the main function \code{BFDA()} by

\code{
> param_rgrid_ns = setOptions_bfda('smethod', 'babf', 'cgrid', 0, 'mat', 0, 'M', 10000, 'Burnin', 2000, 'm', m, 'eval_grid', pgrid, 'ws', 0.05);
}

\code{
> [out_rgrid_ns, param_rgrid_ns] = BFDA(GausFD_rgrid_ns.Xraw_cell, GausFD_rgrid_ns.Tcell, param_rgrid_ns);
}

where \code{mat} is set as $0$ in \code{param_rgrid_ns}.

\subsubsection{Example results}

\begin{figure}[htb]
\begin{center}
\includegraphics[width=0.65\textwidth]{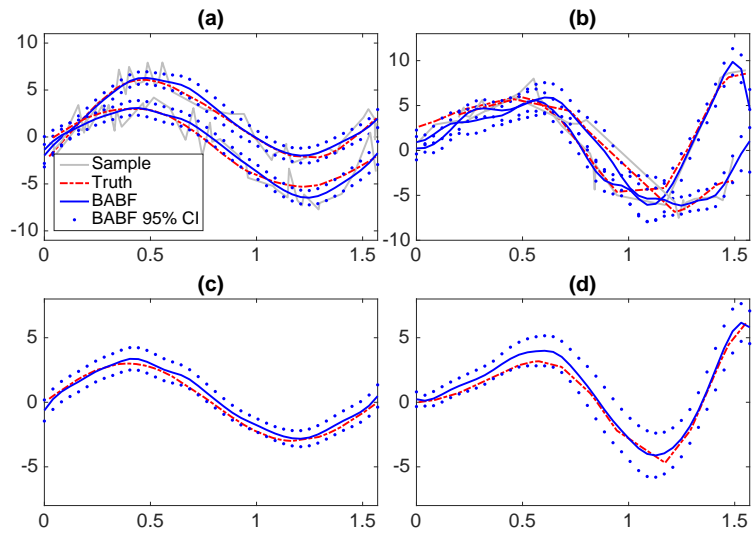}
\caption{Results of analyzing functional data with \textbf{Random grids} by BABF: (a) two sample signal estimates with stationary data; (b) two sample signal estimates with nonstationary data; (c) mean estimate with stationary data; (d) mean estimate with nonstationary data; along with 95\% pointwise credible intervals (blue dots).}
\label{fig:5}
\end{center}
\end{figure}

\begin{figure}[htb]
\begin{center}
\includegraphics[width=0.7\textwidth]{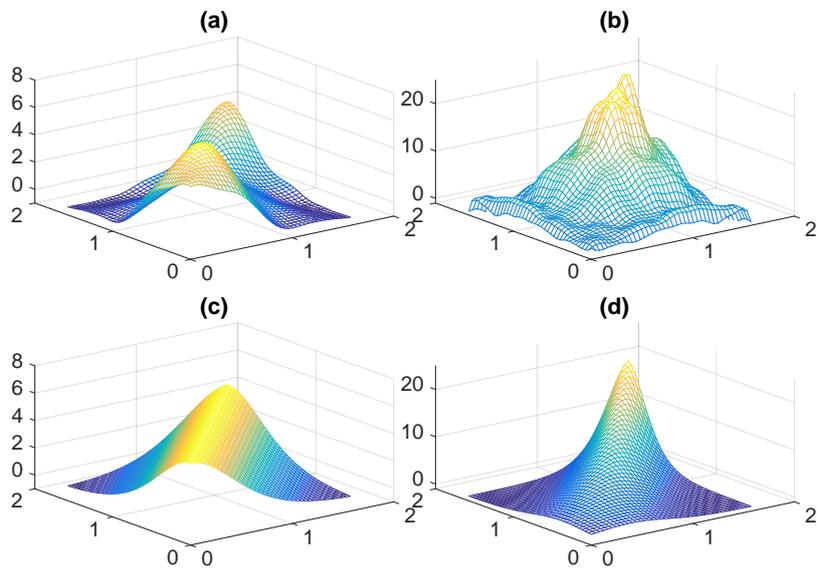}
\caption{Covariance estimates for functional data with \textbf{Random grids}: (a) BABF estimate with stationary data; (b) BABF estimate with nonstationary data; (c) true covariance surface for stationary data; (d) true covariance surface for nonstationary data.}
\label{fig:6}
\end{center}
\end{figure}

With random observation grids, the BABF method can efficiently analyze both stationary and nonstationary functional data, producing smooth estimates for signals and mean-covariance functions that are close to the truth (Figures \ref{fig:5}, \ref{fig:6}). Specifically, the 95\% pointwise credible intervals of signal estimates have coverage probabilities $>0.95$.  

\clearpage

%% file: Regression.tex
\section{Functional linear regression}
\label{reg}
We expect follow-up FDA results will be improved by using the accurately smoothed functional data produced by \pkg{BFDA}. Specifically, we show examples of functional linear regression in this Section, considering the following two models,
\begin{eqnarray}
\bm{Y} &=& \beta_0 + \int \bm{X(t)}^\top \bm{\beta(t)} \; dt + \bm{\epsilon} , \label{vecy} \\
\bm{Y(t)} &=& \beta_0(t) + \bm{X(t)}^\top \bm{\beta(t)}+ \bm{\epsilon(t)} ; \label{fdy}
\end{eqnarray}
where 
\begin{itemize}
\item $\bm{Y}$ in Equation \ref{vecy} denotes a $n\times 1$ vector of scalar responses; $\bm{Y(t)} = (y_1(t), \cdots, y_n(t))^\top$ in Equation \ref{fdy} denotes a vector of functional responses; 
\item  $\bm{X(t)}$ denotes a $n\times q$ design matrix of $q$ functional independent variables; 
\item $\bm{\beta(t)}$ denotes a $q\times 1$ vector of coefficient functions for independent variables; 
\item $\beta_0$ and $\beta_0(t)$ denote the intercept terms; 
\item  $\bm{\epsilon}$ and $\bm{\epsilon(t)}$ denote the error terms. 
\end{itemize}
Note that $\bm{X(t)}$ and $\bm{\beta(t)}$ can also denote nonfunctional covariates and coefficients, because nonfunctional variables can be thought as constant functions of $t$. 

\subsection{Simulate functional data}
To evaluate the improvement of regression results using the smoothed data by \pkg{BFDA}, we first simulated $30$ raw stationary GP trajectories $\{X_i(\bm{t_i})\}$ with random grids uniformly generated from $[0, \pi/2]$, by 
\code{sim_gfd_rgrid(30, 40, 0, 1.5708, 2.2361, 2, 3.5, 0.5, 1)}. Then we simulated the scalar responses by $Y_i = \int_0^{1.5708} X_i(t)t^2 \, dt + \epsilon $, and the functional responses by $Y_i(t) = X_i(t)t^2 + \epsilon$, with $\epsilon \sim N(0, 1)$. 

Because the functional regression function \code{fRegress()} in \pkg{fdaM} requires inputs of functional data with common grids, we interpolated the simulated true data, smoothed data by BABF with \pkg{BFDA}, and the raw data with noises on the equally spaced common grid (with length $40$) over $[0, \pi/2]$, by cubic smoothing spline (CSS, using the function \code{csaps()} with the suggested optimal smoothing parameter $1$). As a result, the interpolated signals from the raw data are equivalent to the individually smoothed ones by CSS (one example curve is shown in Figure \ref{simu_x}(a)).

With the smoothed data by BABF and CSS, we respectively fitted the functional linear models (Equations \ref{vecy} and \ref{fdy}) using $20$ randomly chosen signals, and then tested the prediction results using the remains. We replicated this fitting process for 100 times, and evaluated the performance by the average mean square errors (MSEs) of the fitted and predicted responses.

\subsection{Results with scalar responses}

\begin{figure}[htb]
\begin{center}
\includegraphics[width=0.4\textwidth]{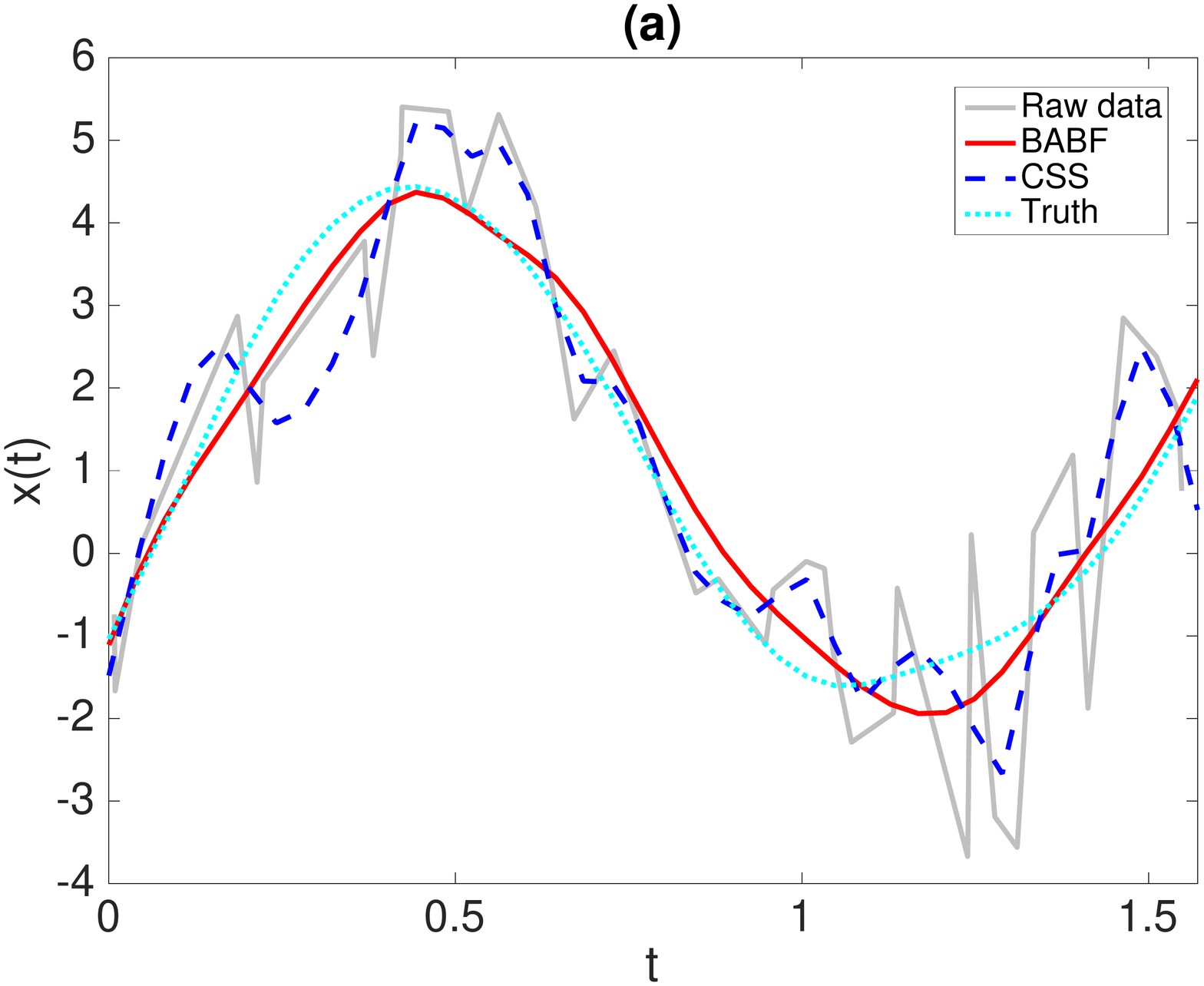}
\includegraphics[width=0.4\textwidth]{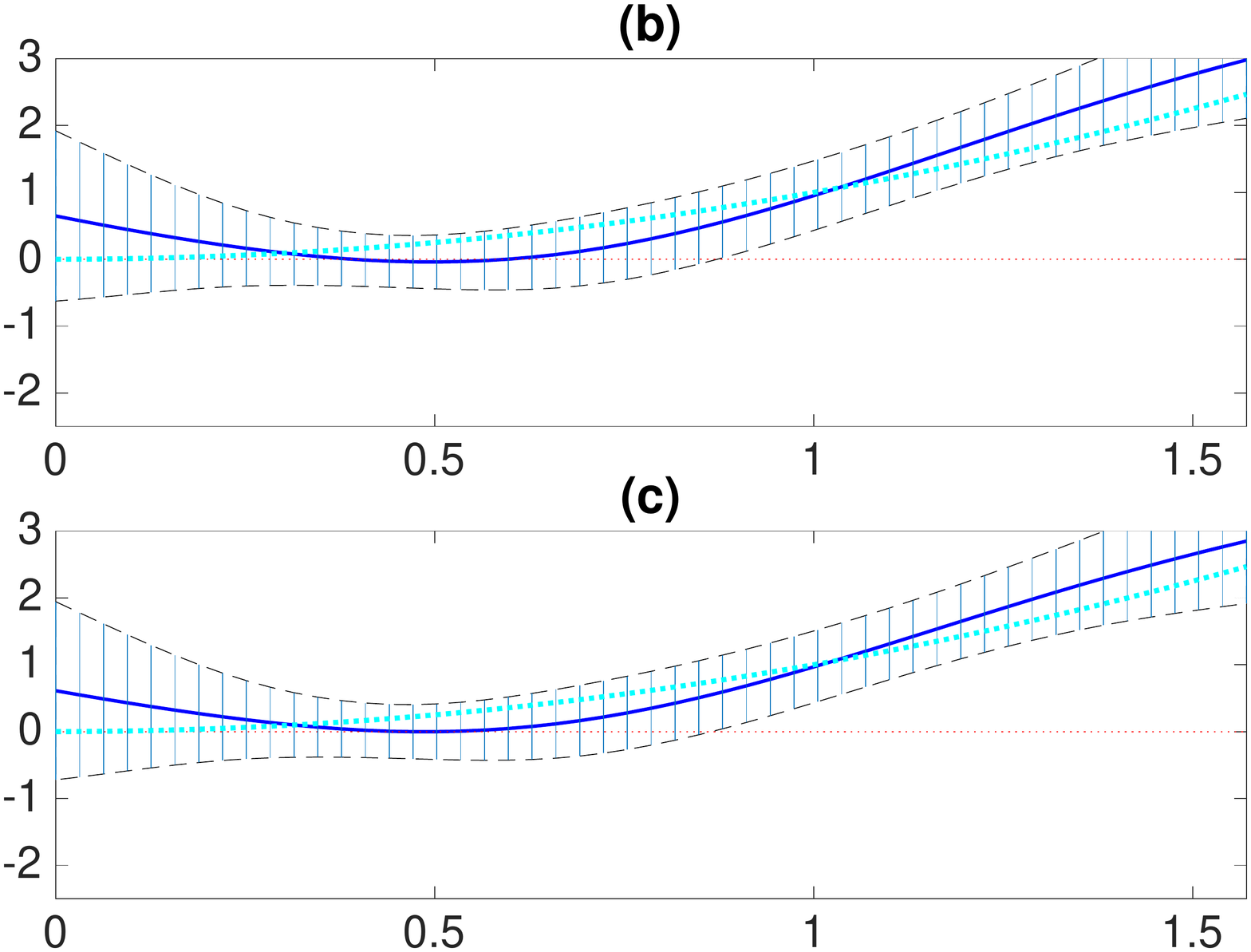}
\caption{(a) Example estimates of $X_i(t)$; (b) the estimate of $\beta(t)$ using the smoothed data by BABF; (c) the estimate of $\beta(t)$ using the smoothed data by CSS; along with the truth in the cyan dotted lines.}
\label{simu_x}
\end{center}
\end{figure}

\begin{figure}[hb]
\begin{center}
\includegraphics[width=0.4\textwidth]{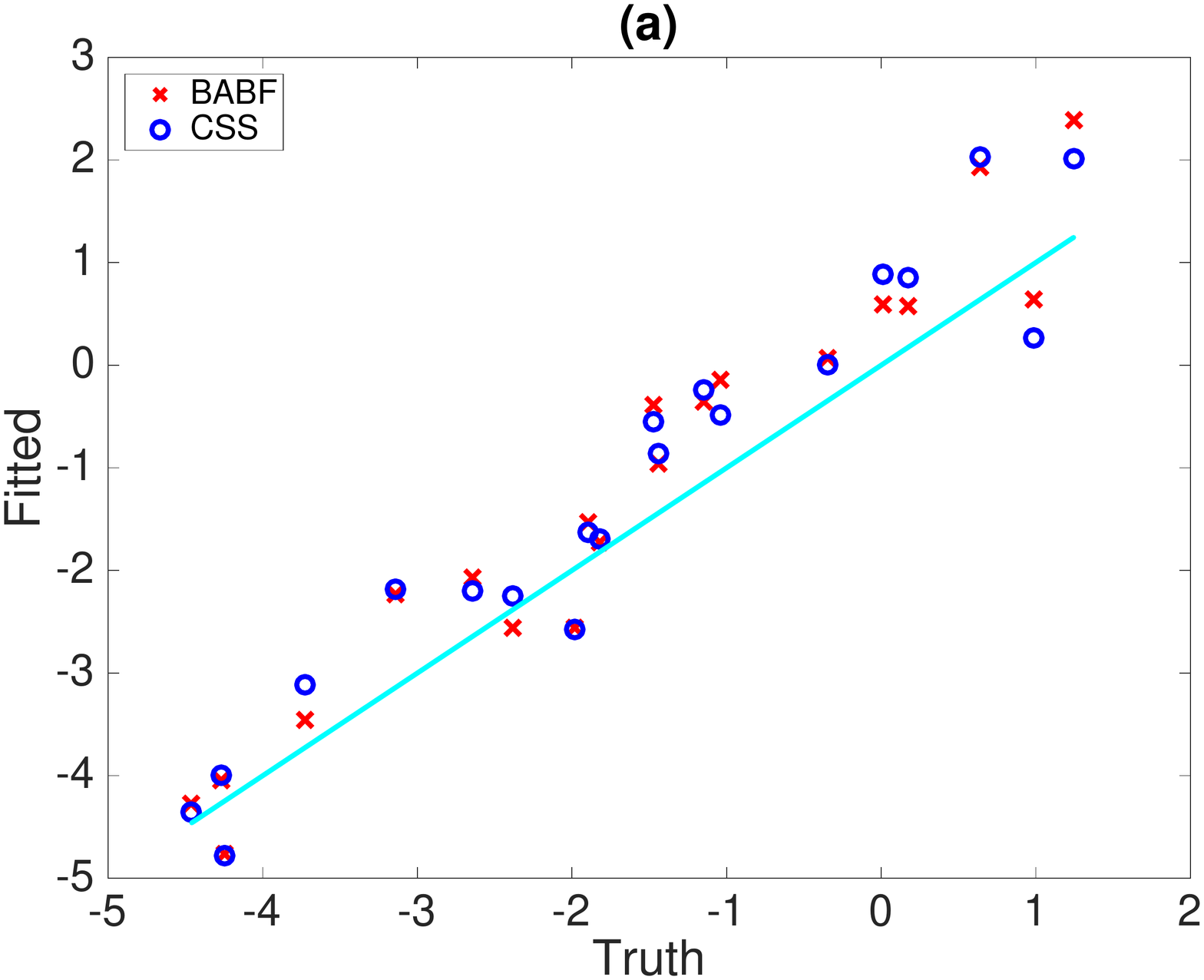}
\includegraphics[width=0.4\textwidth]{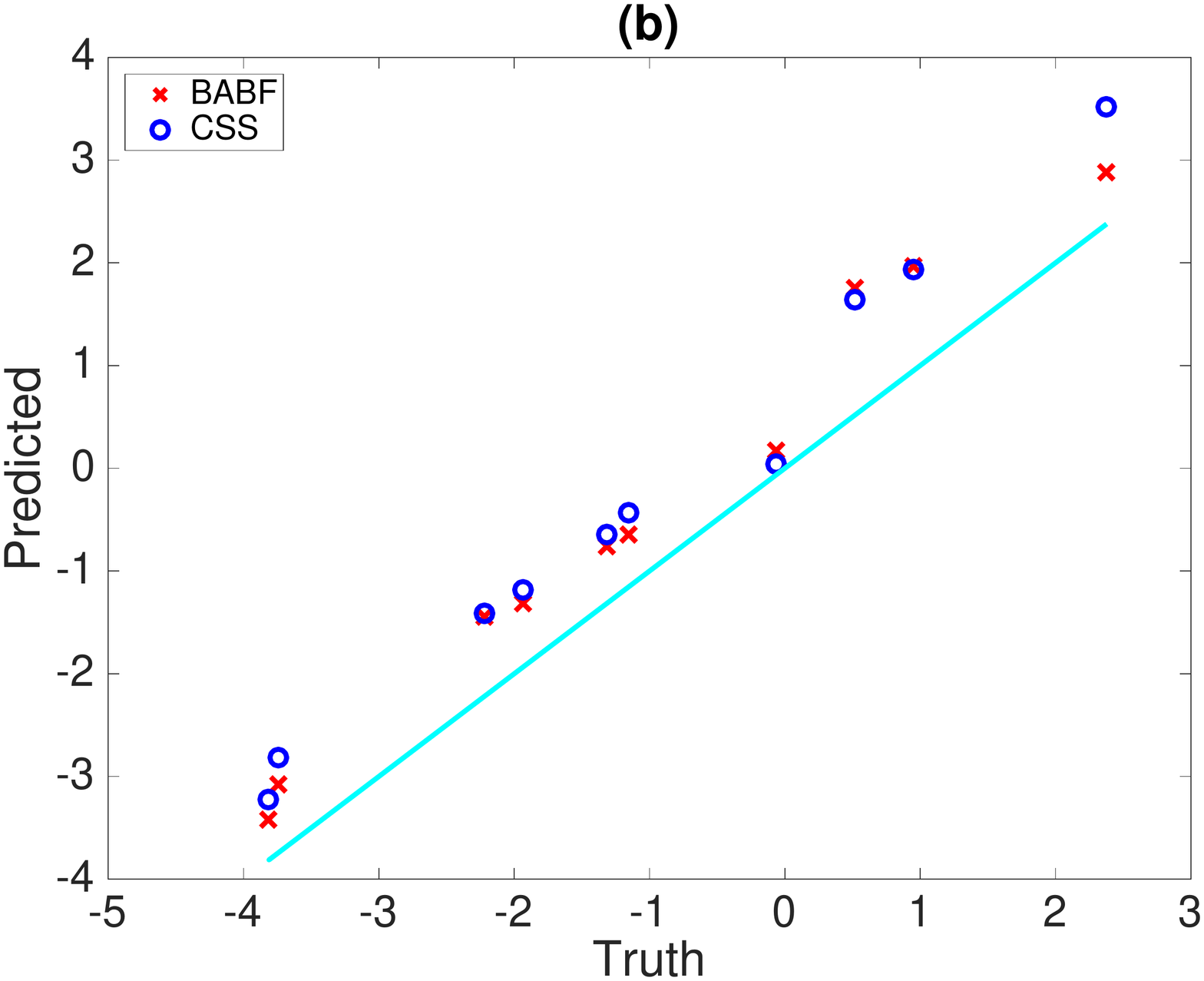}
\caption{(a) Fitted vs.~true scalar responses; (b) predicted vs.~true scalar responses.}
\label{vecy_plot}
\end{center}
\end{figure}

For the case with scalar responses, although the fitted coefficient functions using both smoothed data by BABF and CSS are close to the truth (Figure \ref{simu_x}(b, c)), with coverage probabilities $>0.95$ for the 95\% confidence intervals, the average MSEs of the fitted and predicted responses from 100 replications are smaller for using the BABF smoothed data than the ones using the CSS smoothed data (0.311 vs.~0.388 for fitted responses, 0.497 vs.~0.677 for predicted responses, as shown in Table \ref{tb1}). Figure \ref{vecy_plot} shows the results of an example replication of this fitting and predicting process with scalar responses. 

\subsection{Results with functional responses}

\begin{figure}[htb]
\begin{center}
\includegraphics[width=0.4\textwidth]{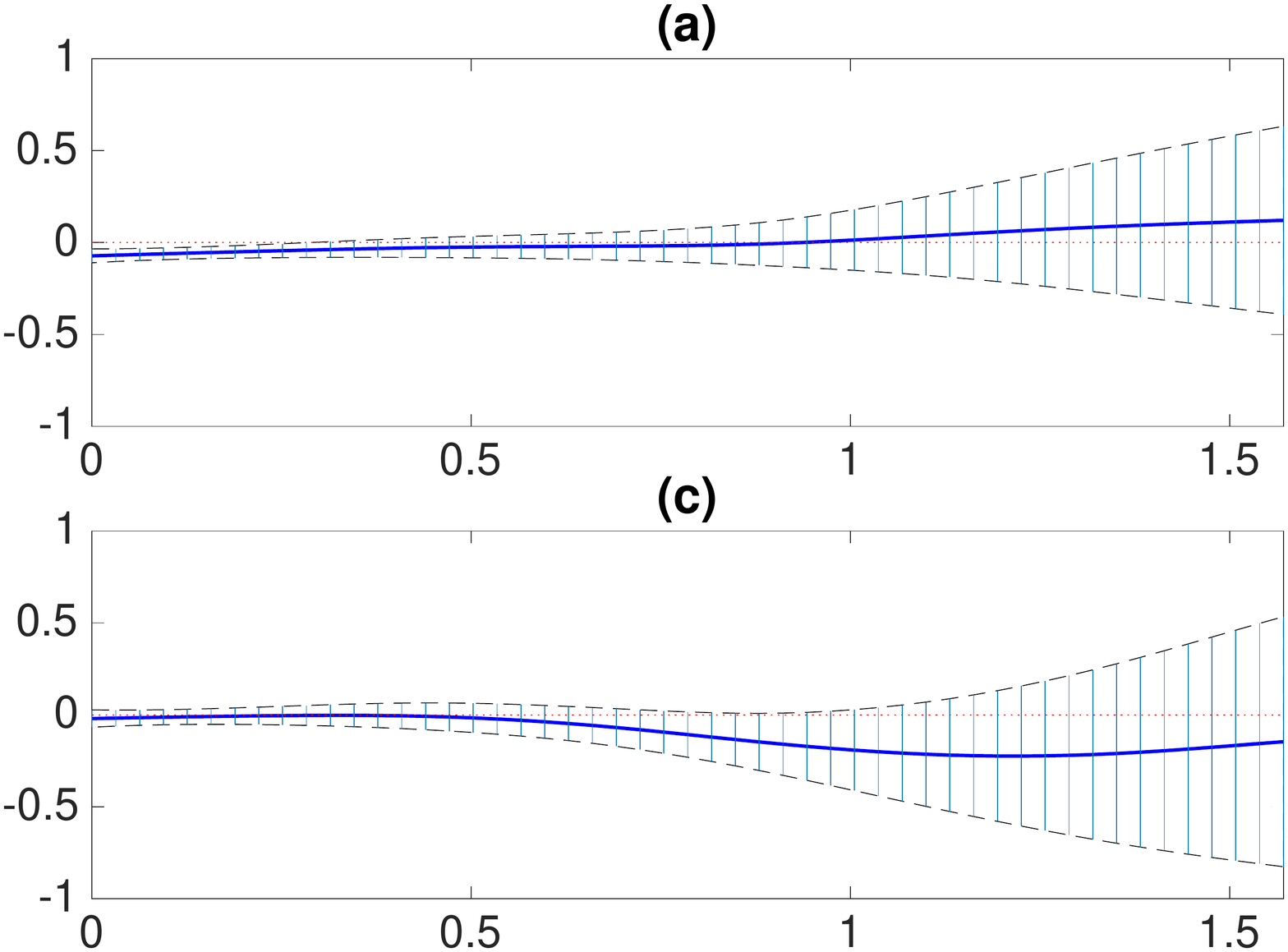}
\includegraphics[width=0.4\textwidth]{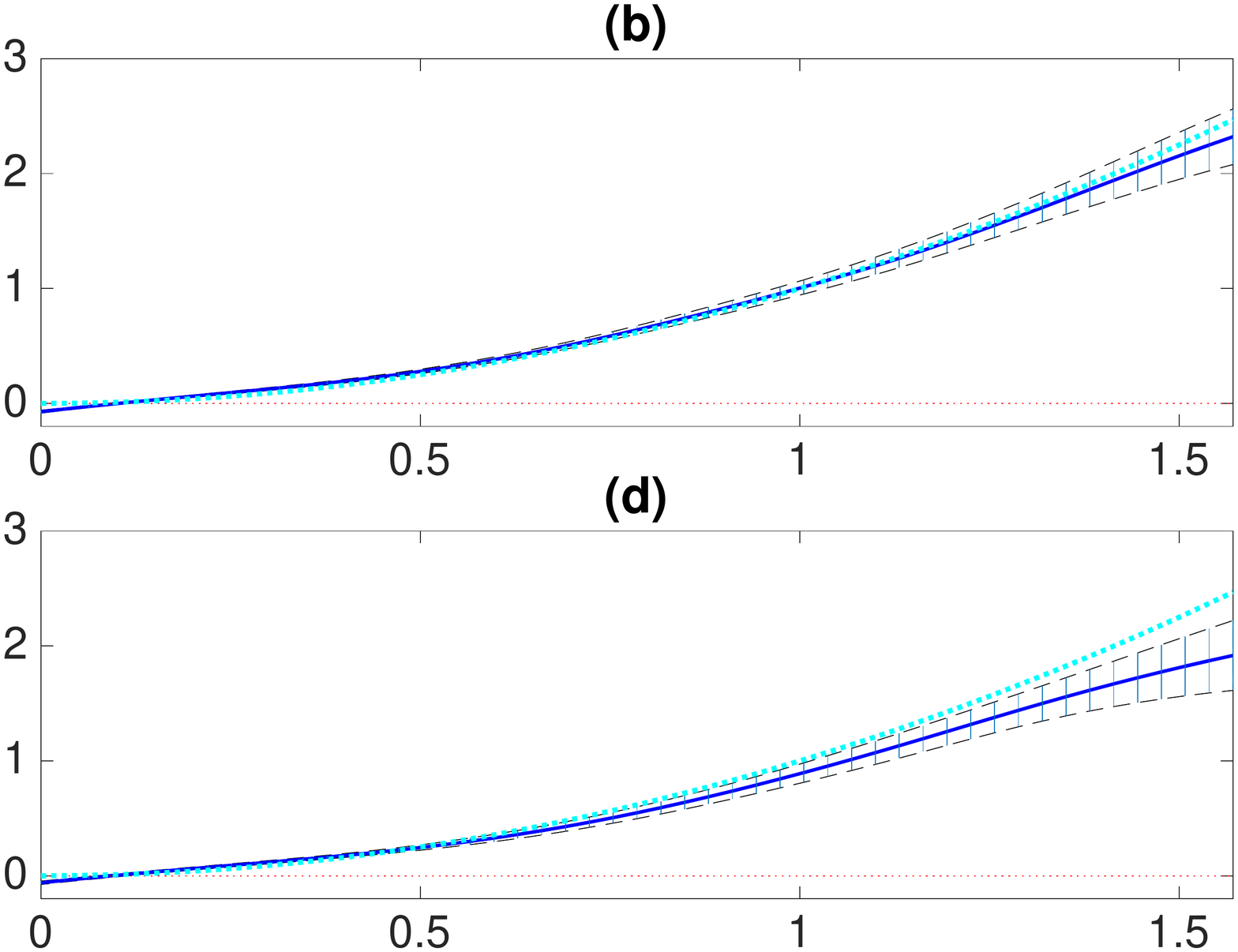}
\caption{(a) Intercept function $\beta_0(t)$ using the BABF smoothed data; (b) Intercept function $\beta_0(t)$ using the CSS smoothed data; (c) coefficient function $\beta(t)$ using the BABF smoothed data; (d) coefficient function $\beta(t)$ using the CSS smoothed data; along with 95\% confidence intervals and true coefficient functions in the cyan dotted lines.}
\label{beta}
\end{center}
\end{figure}

\begin{figure}[htb]
\begin{center}
\includegraphics[width=0.4\textwidth]{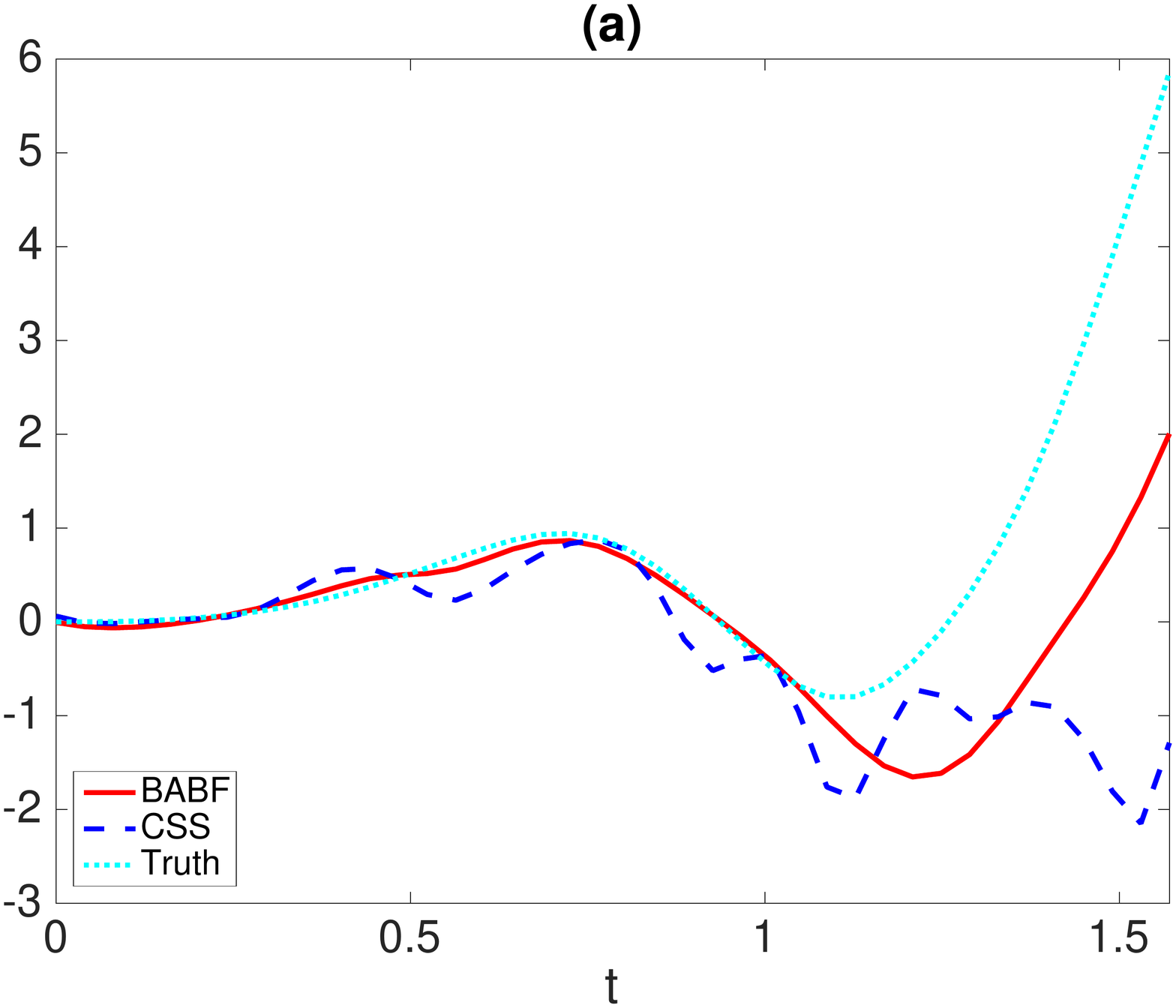}
\includegraphics[width=0.4\textwidth]{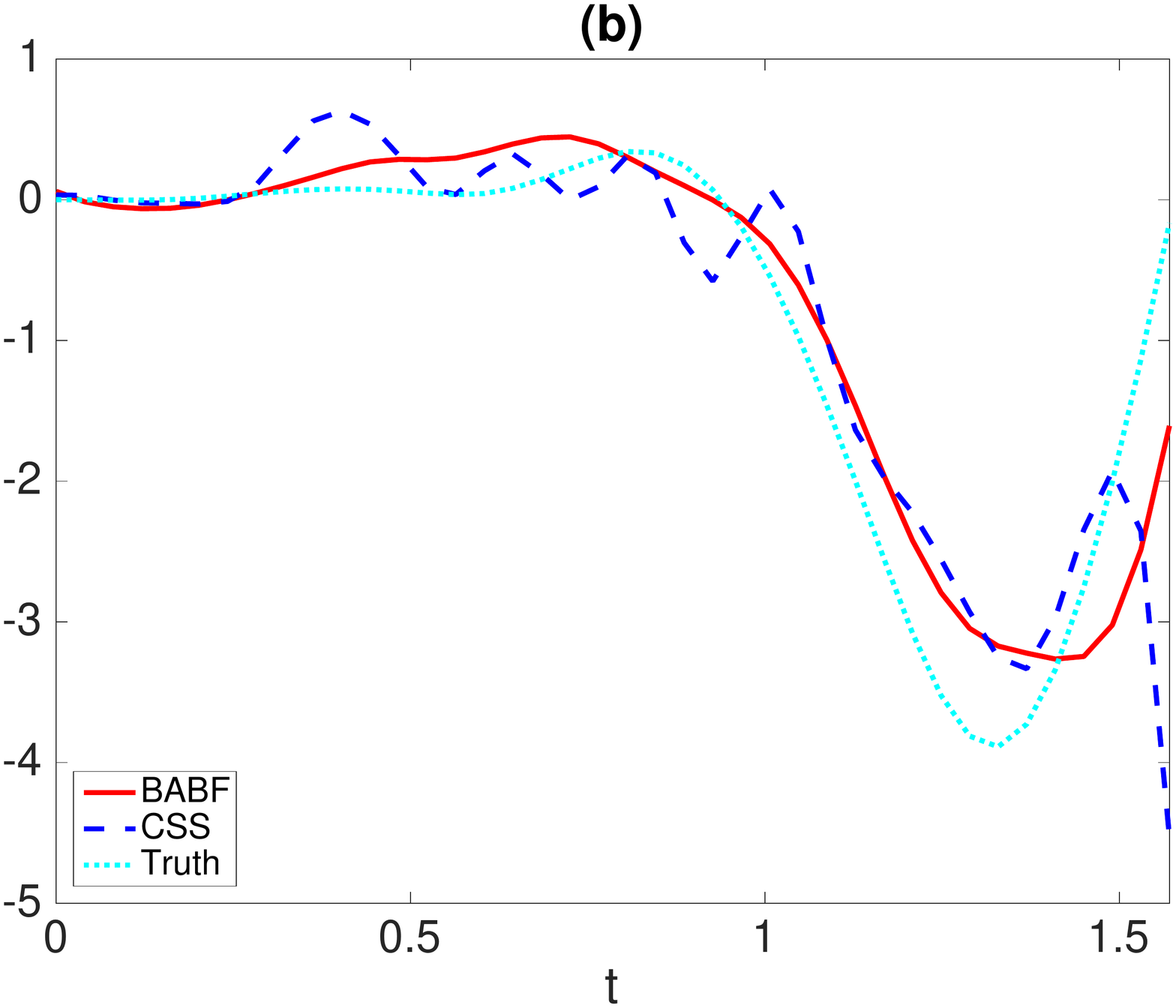}
\caption{(a) Example fitted functional responses; (b) example predicted functional responses; with true signals in the cyan dotted lines.}
\label{fdy_plot}
\end{center}
\end{figure}

For the case with functional responses, we can see that the fitted intercept term using the BABF smoothed data is closer to the truth (constant $0$) with narrower 95\% confidence interval than the one using the CSS smoothed data (Figure \ref{beta}(a, c)). In addition, the coefficient function using BABF smoothed data has narrower 95\% confidence interval but higher coverage probability (Figure \ref{beta}(b, d)). Consequently, both fitted and predicted functional responses using the BABF smoothed data have smaller point-wise MSEs out of 100 replications, 0.417 vs.~1.190 for fitted functional responses, 0.464 vs.~1.354 for predicted functional responses (Table \ref{tb1}). Figure \ref{fdy_plot} shows the results of an example replication of this fitting and predicting process with functional responses.

\begin{table}[htb]
\centering
\begin{tabular}{ c|cc|cc } 
 \hline
\multirow{2}{6em}{\centering MSE (std)}  & \multicolumn{2}{|c|}{BABF smoothed} &  \multicolumn{2}{|c}{CSS smoothed}     \\
		   &$\bm{Y}$ & $\bm{Y(t)}$ & $\bm{Y}$ & $\bm{Y(t)}$ \\
 \hline
Fitted & 0.311 (0.061) & 0.417 (0.049) & 0.388 (0.074) & 1.190 (0.186)  \\ 
Predicted & 0.497 (0.289) & 0.464 (0.112) & 0.677 (0.435) & 1.354 (0.419) \\ 
 \hline
\end{tabular}
 \caption{Average MSEs of the fitted and predicted responses for 100 replicates, along with the standard deviations of these MSEs among 100 replicates in the parentheses, for scalar responses $\bm{Y}$ and functional responses $\bm{Y(t)}$.}
 \label{tb1}
 \end{table}

\clearpage

%% file: Discussion.tex
\section{Discussion}
\label{dis}

The \proglang{MATLAB} tool \pkg{BFDA} presented in this paper can simultaneously smooth multiple functional observations and estimate the mean-covariance functions, assuming the functional data are from the same GP. The smoothed data by \pkg{BFDA} are shown to be more accurate than the conventional individual smoothing methods, thus improving follow-up analysis results. The advantages of \pkg{BFDA} include:
\begin{itemize}
\item Simultaneously smoothing multiple functional samples and estimating mean-covariance functions in a nonparametric way;
\item Flexibly handling functional data with stationary and nonstationary covariance functions, common or uncommon (sparse) observation grids;
\item Efficiently dealing with high-dimensional functional data by the BABF method.
\end{itemize}

\pkg{BFDA} is suitable for analyzing data that can be roughly assumed as from the same GP distribution. We recommend using the BHM method for low-dimensional functional data with common grids or sparse functional data, and using the BABF method for high-dimensional data with dense grids (including both common and random grids). In addition, we recommend using the Mat\'ern function as the prior covariance structure for analyzing functional data with stationary covariance functions, while using the empirical covariance estimate (e.g., the estimate by \pkg{PACE} is recommended) for analyzing functional data with nonstationary covariance functions.

The follow-up functional data analysis can be conducted using the existing softwares (e.g., \pkg{fdaM} in \proglang{MATLAB}, \pkg{fda} in \proglang{R}). Examples are provided in \pkg{BFDA} about using \pkg{fdaM} with the smoothed data by \pkg{BFDA}, showing  improved regression results than using the individually smoothed data. Details about the inputs and outputs of \pkg{BFDA}, and part of the example \proglang{MATLAB} scripts are provided in the Appendices. The \pkg{BFDA} tool and example scripts are freely available at \url{https://github.com/yjingj/BFDA}. We will continue integrating more options of basis functions, Bayesian functional data regression using GPs, and functional classification into \pkg{BFDA}. 



%% file: Appendix.tex
\begin{appendices}

\section{Inputs and outputs}

\subsection{Input variables}
The main function \code{BFDA()} has three input arguments:
\begin{itemize}
\item A cell contains all functional data;
\item A cell contains all grids on which the functional data are collected;
\item A parameter structure outputted by function \code{setOptions_bfda()}, containing all required parameters:
\begin{itemize}
\item \code{smethod}, specifying the method used for analyzing the functional data. Default value is \code{'babf'} for BABF method with basis function approximation; other choices are \code{'bhm'} for BHM method without basis function approximation, \code{'bgp'} for standard Bayesian GP regression, \code{'bfpca'} for Bayesian principal components analysis;
\item \code{Burnin}, the number of burn-ins for the MCMC algorithm. Default value is \code{2000};
\item \code{M}, the number of iterations for the MCMC algorithm. Default value is \code{10000};
\item \code{cgrid}, set as \code{1} if the functional data are observed on a common-grid, otherwise set as \code{0} for uncommon or random grids. Default value is \code{1};
\item \code{Sigma_est}, estimated smooth covariance matrix from previous analysis. Default is empty and will be estimated by \pkg{PACE} or sample estimate from individually smoothed data;
\item \code{mu_est}, estimated smooth functional mean from previous analysis. Default is empty and will be set as the smoothed sample mean;
\item \code{mat}, set as \code{1} to use the Mate\'rn covariance function as prior structure for stationary functional data; set as \code{0} to use the empirical covariance estimate \code{Sigma_est} as the prior structure for nonstationary functional data. Default value is \code{1};
\item \code{nu}, order of smoothness for the Mate\'rn covariance function. Default is empty and will be estimated based on \code{Sigma_est};
\item \code{delta}, shape parameter $\delta$ of the IWP. Default is \code{5} for a non-informative prior;
\item \code{c}, determining the prior covariance for functional mean. Default is \code{1};
\item \code{w, ws}, determining the prior gamma distributions for $\sigma^2_{\epsilon}$ and $\sigma^2_s$. Defaults are \code{w}=1, \code{ws}=0.1. The parameter \code{ws} should be tuned for a proper magnitude of the posterior covariance estimate; 
\item \code{pace}, if \code{Sigma_est} and \code{mu_est} are empty, set \code{pace}=1 to obtain \code{Sigma_est} and \code{mu_est} by \pkg{PACE}, and set \code{pace}=0 to use the empirical estimates from the individually smoothed data by CSS. Default is \code{1};
\item \code{m, tau},  working grid \code{tau} is only required for \code{'babf'} method. Default is empty and will be set up as the $(0:\frac{100}{m-1}:100)$ percentiles of the pooled observation grid with length \code{m};
\item \code{eval_grid}, evaluation grid for all functional estimates, only required for \code{'babf'} methods;
\item \code{lamb_min, lamb_max, lamb_step}, determining the smoothing parameter candidates for general cross validation of the CSS method. Defaults are \code{lamb_min}=0.9, \code{lamb_max}=0.99, \code{lamb_step}=0.01;
\item \code{a, b}, hyper parameters for the gamma distributions in \code{'bgp'}, and \code{'bfpca'}.
\item \code{resid_thin}, determine the MCMC thinning steps of the residuals that are used to test the goodness-of-fit of the model. Default is \code{10}.  
\end{itemize}
\end{itemize} 

\subsection{Output variables}
The main function \code{BFDA()} has two output arguments, one structure outputted by the specified method, and the other parameter structure as specified by \code{setOptions_bfda()} containing updated parameter values.

Output structure with \code{smethod} = \code{'bhm'}:
\begin{itemize}
\item \code{Z, Z_CL, Z_UL}, smoothed functional data, lower and upper 95\% credible intervals;
\item \code{Sigma, Sigma_CL, Sigma_UL}, functional covariance estimate, lower and upper 95\% credible intervals;
\item \code{Sigma_SE}, the empirical covariance estimate by using the smoothed data \code{Z};
\item \code{mu, mu_CI}, functional mean estimate, 95\% credible intervals;
\item  \code{rn, rn_CI}, estimate and 95\% credible interval for the noise precision;
\item  \code{rs, rs_CI}, estimate and 95\% credible interval for $\sigma^2_s$;
\item \code{rho, nu}, estimated parameter values for the Mat\'ern function;
\item \code{residuals}, MCMC samples of the residuals that are used to test the goodness-of-fit;
\item \code{pmin_vec}, p-values for testing the goodness-of-fit for all functional samples. P-value > 0.25 suggests no evidence of model inadequacy; 0.05 < p-value < 0.25 suggests some evidence of model inadequacy; p-value < 0.05 suggests strong evidence of model inadequacy.
\end{itemize}

The output structure with \code{smethod} = \code{'babf'} has the following variables that are different from the ones with \code{smethod} = \code{'bhm'}:
\begin{itemize}
\item \code{Zt}, smoothed functional data on the observation grids;
\item \code{Z_cgrid, Z_cgrid_CL, Z_cgrid_UL}, smoothed functional data on the evaluation grid \code{eval_grid}, along with lower and upper 95\% credible intervals;
\item \code{Sigma_cgrid, Sigma_cgrid_CL, Sigma_cgrid_UL}, functional covariance estimate on the evaluation grid \code{eval_grid}, along with lower and upper 95\% credible intervals;
\item \code{mu_cgrid, mu_cgrid_CI}, functional mean estimate on the evaluation grid \code{eval_grid}, along with 95\% credible intervals;

\item \code{Zeta, Zeta_CL, Zeta_UL}, estimates for the coefficients of basis functions, along with lower and upper 95\% credible intervals;
\item \code{Sigma_zeta_SE}, the empirical covariance estimate with the estimated \code{Zeta};
\item \code{Sigma_zeta, Sigma_zeta_CL, Sigma_zeta_UL}, covariance estimate for the coefficients of basis functions, along with lower and upper 95\% credible intervals;
\item \code{mu_zeta, mu_zeta_CI}, mean estimate for the coefficients of basis functions, along with 95\% credible intervals;
\item \code{Btau}, the basis function evaluations on the working grid \code{tau};
\item \code{BT}, the basis function evaluations on the observation grids;

\item \code{Sigma_tau}, functional covariance estimate on the working grid \code{tau};
\item \code{mu_tau}, functional mean estimate on the working grid \code{tau};
\item \code{optknots}, the optimal knots selected by \code{optknt()} for evaluations on the working grid \code{tau}.

\end{itemize}

\section[Example MATLAB scripts for using BFDA]{Example \proglang{MATLAB} scripts for using \pkg{BFDA}}

\begin{verbatim}
%% -------- Add pathes of the required MATLAB packages -------- 
% BFDA, bspline, fdaM, mcmcdiag, PACE
% replace pwd by the directory of your MATLAB packages
addpath(genpath(cat(2, pwd, '/BFDA')))
addpath(genpath(cat(2, pwd, '/bspline')))
addpath(genpath(cat(2, pwd, '/fdaM'))) 
addpath(genpath(cat(2, pwd, '/mcmcdiag'))) 
addpath(genpath(cat(2, pwd, '/PACErelease2.11')))

%% -------- Set up parameters for simulation -------- 
n = 30; % Number of functional samples
p = 40; % Number of pooled grid points, or evaluated grid points
s = sqrt(5); % Standard deviation of functional observations
r = 2; % Signal to noise ratio
rho = 1/2; % Scale parameter in the Matern function
nu = 3.5; % Order parameter in the Matern function
pgrid = (0 : (pi/2)/(p-1) : (pi/2)); % Pooled grid
dense = 0.6; % Proportion of observations on the pooled grid
au = 0; bu = pi/2; % Function domain 
m = 20; % Number of working grid points
stat = 1; % Specify stationary data
cgrid = 1; % Specify common observation grid

%% -------- Analyzing stationary functional data with common grid -------- 
% Generate simulated data from GP(3sin(4t), s^2 Matern_cor(d; rho, nu)) 
% with noises from N(0, (s/r)^2)
GausFD_cgrid = sim_gfd(pgrid, n, s, r, nu, rho, dense, cgrid, stat);

% setup parameters for BFDA
% run with BHM
param = setOptions_bfda('smethod', 'bhm', 'cgrid', 1, 'mat', 1, ...
           'M', 10000, 'Burnin', 2000, 'w', 1, 'ws', 1);

% run with Bayesian Functional PCA
% param = setOptions_bfda('smethod', 'bfpca', 'M', 50, 'Burnin', 20);

% run with standard Bayesian Gaussian Process model
% param = setOptions_bfda('smethod', 'bgp', 'mat', 1, ...
%           'M', 50, 'Burnin', 20);

% run with Cubic Smoothing Splines
% param = setOptions_bfda('smethod', 'css', 'mat', 1, 'M', ...
%            50, 'Burnin', 20, 'pace', 0);

% call BFDA
[out_cgrid, param ] = ...
    BFDA(GausFD_cgrid.Xraw_cell, GausFD_cgrid.Tcell, param);

%% -------- Analyzing stationary functional data with uncommon grid -------- 
GausFD_ucgrid = sim_gfd(pgrid, n, s, r, nu, rho, dense, 0, stat);

param_uc = setOptions_bfda('smethod', 'bhm', 'cgrid', 0, 'mat', 1, 'M',...
    10000, 'Burnin', 2000, 'pace', 1, 'ws', 0.1);

[out_ucgrid, param_uc] = ...
    BFDA(GausFD_ucgrid.Xraw_cell, GausFD_ucgrid.Tcell, param_uc);

%% -------- Analyzing non-stationary functional data with common grid -------- 
GausFD_cgrid_ns = sim_gfd(pgrid, n, s, r, nu, rho, dense, cgrid, 0);

param_ns = setOptions_bfda('smethod', 'bhm', 'cgrid', 1, 'mat', 0, 'M',...
    10000, 'Burnin', 2000, 'pace', 1, 'ws', 0.01);

[out_cgrid_ns, param_ns] = ...
    BFDA(GausFD_cgrid_ns.Xraw_cell, GausFD_cgrid_ns.Tcell, param_ns);

%% --------  Analyzing non-stationary functional data with uncommon grid -------- 
GausFD_ucgrid_ns = sim_gfd(pgrid, n, s, r, nu, rho, dense, 0, 0);

param_uc_ns = setOptions_bfda('smethod', 'bhm', 'cgrid', 0, 'mat', 0, ...
    'M', 10000, 'Burnin', 2000, 'pace', 1, 'ws', 0.01);

[out_ucgrid_ns, param_uc_ns ] = ...
    BFDA(GausFD_ucgrid_ns.Xraw_cell, GausFD_ucgrid_ns.Tcell, param_uc_ns);

%% -------- Analyzing stationary functional data with random grids -------- 
GausFD_rgrid = sim_gfd_rgrid(n, p, au, bu, s, r, nu, rho, stat);

param_rgrid = setOptions_bfda('smethod', 'babf', 'cgrid', 0, 'mat', 1, ...
    'M', 10000, 'Burnin', 2000, 'm', m, 'eval_grid', pgrid, 'ws', 1, ...
    'trange', [au, bu]);

% call BFDA
[out_rgrid, param_rgrid]= ...
    BFDA(GausFD_rgrid.Xraw_cell, GausFD_rgrid.Tcell, param_rgrid);

%% -------- Analyzing nonstationary functional data with random grids-------- 
GausFD_rgrid_ns = sim_gfd_rgrid(n, p, au, bu, s, r, nu, rho, 0);

param_rgrid_ns = setOptions_bfda('smethod', 'babf', 'cgrid', 0, 'mat', ...
    0, 'M', 10000, 'Burnin', 2000, 'm', m, 'eval_grid', pgrid, 'ws', 0.05, ...
    'trange', [au, bu]);

% call BFDA
[out_rgrid_ns, param_rgrid_ns] = ...
    BFDA(GausFD_rgrid_ns.Xraw_cell, GausFD_rgrid_ns.Tcell, param_rgrid_ns);

%% -------- Calculate RMSE (root mean square error) -------- 
display('RMSE of the estimated stationary covariance')
rmse(out_cgrid.Sigma_SE, GausFD_cgrid.Cov_true)

display('RMSE of the estimated functional data')
Xtrue_mat = reshape(cell2mat(GausFD_cgrid.Xtrue_cell), p, n);
rmse(out_cgrid.Z, Xtrue_mat)

display('RMSE of the estimated non-stationary covariance')
Ctrue_ns = cov_ns(pgrid, sf, nu, rho); 
% calculate the true non-stationary covariance matrix
rmse(out_cgrid_ns.Sigma_SE, Ctrue_ns)

%% --------  Save simulated data sets and BFDA results -------- 
save('./Examples/Data/Simu_Data.mat', 'GausFD_cgrid', 'GausFD_ucgrid', ...
                           'GausFD_cgrid_ns', 'GausFD_ucgrid_ns', ...
                           'GausFD_rgrid', 'GausFD_rgrid_ns')

save('./Examples/Data/Simu_Output.mat', 'out_cgrid', 'out_ucgrid', ...
                           'out_cgrid_ns', 'out_ucgrid_ns', ...
                           'out_rgrid', 'out_rgrid_ns')
\end{verbatim}


\section[Example MATLAB scripts for functional regression using fdaM]{Example \proglang{MATLAB} scripts for functional regression using \pkg{fdaM}}

\begin{verbatim}
%% -------- Add fdaM path and load the functional data -------- 
% Replace pwd by the directory of your MATLAB packages
addpath(genpath(cat(2, pwd, '/fdaM')));
load('./Examples/Data/Simu_Data.mat');
load('./Examples/Data/Simu_Output.mat');

%% -------- Set sample sizes, training data set, and test data set -------- 
n = 30; % Number of functional curves
p = 40; % Number of pooled grid points, or evaluated grid points
au = 0; bu = pi/2; % Domain of t
pgrid = (au : (bu)/(p-1) : bu); % Pooled grid
trange = [au, bu];

sampind = sort(randsample(1:n,20,false)) ;
samptest = find(~ismember(1:n, sampind));
n_train = length(sampind); n_test = length(samptest);

cgrid = 0;

Xtrue = zeros(p, n);
Xraw = zeros(p, n);
Xsmooth = zeros(p, n);

if cgrid
% Functional observations with common grids
    Xtrue = reshape(cell2mat(GausFD_cgrid.Xtrue_cell), p, n);
    Xsmooth = out_cgrid.Z(:, sampind);
    Xraw = reshape(cell2mat(GausFD_cgrid.Xraw_cell), p, n);
else
% Functional observations with random grids
    for i = 1:n
        xi = GausFD_rgrid.Xtrue_cell{i};
        xrawi = GausFD_rgrid.Xraw_cell{i};
        ti = GausFD_rgrid.Tcell{i};
        %interpolate by cubic smoothing spline
        h = mean(diff(ti));
        Xtrue(:, i) = csaps(ti, xi, 1/(1 + h^3/6), pgrid); 
        Xraw(:, i) = csaps(ti, xrawi, 1/(1 + h^3/6), pgrid);
        zi = out_rgrid.Zt{i};
        Xsmooth(:, i) = csaps(ti, zi, 1/(1 + h^3/6), pgrid);
    end
    % Xsmooth = out_rgrid.Z_cgrid;
    % Xsmooth(1, :) = Xsmooth(2, :) * 0.9; 
end

Xtrain = Xsmooth(:, sampind);
Xtest = Xsmooth(:, samptest);
Xraw_train = Xraw(:, sampind);
Xraw_test = Xraw(:, samptest);

rmse(Xtrue, Xsmooth)
rmse(Xtrue, Xraw)

%% --------  Generate response variables -------- 
betamat = (pgrid') .^ 2 ;

% --------  Scalar respones
deltat  = pgrid(2)-pgrid(1);
Avec_true = deltat.*(Xtrue'*betamat    - ...
          0.5.*(Xtrue(1,    :)'.*betamat(1) + Xtrue(p,:)'.*betamat(p)) );
Avec = Avec_true + normrnd(0, 1, n, 1);

Avec_train = Avec(sampind);
Avec_test = Avec(samptest);
Avec_train_true = Avec_true(sampind);
Avec_test_true = Avec_true(samptest);

% --------  Functional responses
ymat_true = Xtrue .* repmat(betamat, 1, n) ;
ymat = ymat_true + normrnd(0, 1, p, n);

ymat_train = ymat(:, sampind);
ymat_test = ymat(:, samptest);
ymat_train_true = ymat_true(:, sampind);
ymat_test_true = ymat_true(:, samptest);

%% -------- Set up functional data structure xfd, yfd for fdaM ----
xnbasis = 20;
xbasis = create_bspline_basis(trange, xnbasis, 4);

xfd_true = smooth_basis(pgrid, Xtrue, xbasis);
xfd = smooth_basis(pgrid, Xtrain, xbasis);
xfd_raw = smooth_basis(pgrid, Xraw_train, xbasis);
[yfd_samp, df, gcv, beta, SSE, penmat, y2cMap, argvals, y] = ...
    smooth_basis(pgrid, ymat_train, xbasis);

%% -------- Set up the curvature penalty operator  -------
conbasis = create_constant_basis(trange); %  create a constant basis
wfd = fd([0, 1], conbasis);
wfdcell = fd2cell(wfd);
curvLfd = Lfd(2, wfdcell);

% Set up xfdcell
xfdcell = cell(1, 2);
xfdcell{1} = fd(ones(1, n_train), conbasis);
xfdcell{2} = xfd;

xfd_raw_cell = cell(1, 2);
xfd_raw_cell{1} = fd(ones(1, n_train), conbasis);
xfd_raw_cell{2} = xfd_raw;

% Set up betacell for scalar responses
betafd0 = fd(0, conbasis);
bnbasis = 10;
betabasis = create_bspline_basis(trange, bnbasis, 4);
betafd1 = fd(zeros(bnbasis, 1), betabasis);

betacell_vecy = cell(1, 2);
betacell_vecy{1} = fdPar(betafd0);
betacell_vecy{2} = fdPar(betafd1, curvLfd, 0);

% Set up betacell, yfd_par for functional responses
betacell_fdy = cell(1, 2);
betacell_fdy{1} = fdPar(betafd1, curvLfd, 0);
betacell_fdy{2} = fdPar(betafd1, curvLfd, 0);

yfd_par = fdPar(yfd_samp, curvLfd, 0);

%% ---- Compute cross-validated SSE's for a range of smoothing parameters ----
%{
wt = ones(1, length(sampind));
lam = (0:0.1:1);
nlam   = length(lam);

SSE_CV_vecy = zeros(nlam,1);
SSE_CV_raw_vecy = zeros(nlam, 1);

SSE_CV_fdy = zeros(nlam,1);
SSE_CV_raw_fdy = zeros(nlam, 1);

for ilam = 1:nlam;
   lambda_vecy       = lam(ilam);
    betacelli_vecy    = betacell_vecy;
    betacelli_vecy{2} = putlambda(betacell_vecy{2}, lambda_vecy);
    SSE_CV_vecy(ilam) = fRegress_CV(Avec_train, xfdcell, betacelli_vecy, wt);
    fprintf('Scalar responses, lambda %6.2f: SSE = %10.4f\n', ...
                lam(ilam), SSE_CV_vecy(ilam));
    
    SSE_CV_raw_vecy(ilam) = fRegress_CV(Avec_train, xfd_raw_cell, betacelli_vecy, wt);
    fprintf('Scalar responses, lambda %6.2f: SSE = %10.4f\n', ...
                lam(ilam), SSE_CV_raw_vecy(ilam));
    
    betacelli_fdy    = betacell_fdy;
    betacelli_fdy{1} = putlambda(betacell_fdy{1}, lambda_vecy);
    betacelli_fdy{2} = putlambda(betacell_fdy{2}, lambda_vecy);
    yfd_par_i = putlambda(yfd_par, lambda_vecy);
    
    SSE_CV_fdy(ilam) = fRegress_CV(yfd_par_i, xfdcell, betacelli_fdy, wt);
    fprintf('Functional respones, lambda %6.2f: SSE = %10.4f\n', ...
                lam(ilam), SSE_CV_fdy(ilam));
    
    SSE_CV_raw_fdy(ilam) = fRegress_CV(yfd_par_i, xfd_raw_cell, betacelli_fdy, wt);
    fprintf('Functional respones, lambda %6.2f: SSE = %10.4f\n', ...
                lam(ilam), SSE_CV_raw_fdy(ilam));
end
%}

%% -------- Fit the linear model -------- 
lambda = 0.1;
wt = ones(1, length(sampind));

% --------- Scalar responses
betacell_vecy{2} = fdPar(betafd1, curvLfd, lambda);

fRegressStruct_vecy = fRegress(Avec_train, xfdcell, betacell_vecy, wt);
fRegressStruct_raw_vecy = ...
    fRegress(Avec_train, xfd_raw_cell, betacell_vecy, wt);

% Get coefficients
betaestcell_vecy   = fRegressStruct_vecy.betahat; 
Avec_hat = fRegressStruct_vecy.yhat;
intercept_vecy = getcoef(getfd(betaestcell_vecy{1}));
disp(['Constant term = ',num2str(intercept_vecy)])

betaestcell_raw_vecy   = fRegressStruct_raw_vecy.betahat; 
Avec_hat_raw = fRegressStruct_raw_vecy.yhat;
intercept_raw_vecy = getcoef(getfd(betaestcell_raw_vecy{1}));
disp(['Constant term = ',num2str(intercept_raw_vecy)])

display(['Scalar reponses:', 'fitted mse = ', ...
    num2str(mse(Avec_train_true, Avec_hat)), ...
    '; fitted mse_raw = ',num2str(mse(Avec_train_true, Avec_hat_raw))])

% Compute Rsquare
covmat = cov([Avec_train, Avec_hat]);
Rsqrd = covmat(1,2)^2/(covmat(1,1)*covmat(2,2));
disp(['R-squared = ',num2str(Rsqrd)])

covmat_raw = cov([Avec_train, Avec_hat_raw]);
Rsqrd_raw = covmat_raw(1,2)^2/(covmat_raw(1,1)*covmat_raw(2,2));
disp(['raw R-squared = ',num2str(Rsqrd_raw)])

% Compute sigma
resid_vecy = Avec_train - Avec_hat;
SigmaE_vecy = mean(resid_vecy.^2);
disp(['Scalar responses: SigmaE = ',num2str(SigmaE_vecy)])

resid_raw_vecy = Avec_train - Avec_hat_raw;
SigmaE_raw_vecy = mean(resid_raw_vecy.^2);
disp(['Scalar responses: Raw SigmaE = ',num2str(SigmaE_raw_vecy)])

% ---------- Functional responses
betacell_fdy{1} = fdPar(betafd1, curvLfd, lambda);
betacell_fdy{2} = fdPar(betafd1, curvLfd, lambda);
yfd_par = fdPar(yfd_samp, curvLfd, lambda);

fRegressStruct_fdy = fRegress(yfd_par, xfdcell, betacell_fdy, wt, y2cMap);
fRegressStruct_raw_fdy = ...
    fRegress(yfd_par, xfd_raw_cell, betacell_fdy, wt, y2cMap);

betaestcell_fdy   = fRegressStruct_fdy.betahat; 
yfd_hat = fRegressStruct_fdy.yhat;
intercept_fdy = eval_fd(pgrid, getfd(betaestcell_fdy{1}));

betaestcell_raw_fdy   = fRegressStruct_raw_fdy.betahat; 
yfd_hat_raw = fRegressStruct_raw_fdy.yhat;
intercept_raw_fdy = eval_fd(pgrid, getfd(betaestcell_raw_fdy{1}));

% MSE of fitted responses 
ymat_fitted = eval_fd(pgrid, yfd_hat);
ymat_fitted_raw = eval_fd(pgrid, yfd_hat_raw);

display(['mse = ', num2str(mse(ymat_train_true, ymat_fitted)), ...
    '; mse_raw = ',num2str(mse(ymat_train_true, ymat_fitted_raw))])
 
% Compute squared residual correlation
resid_fdy = ymat_train_true - ymat_fitted;
SigmaE_fdy = cov(resid_fdy');

resid_raw_fdy = ymat_train_true - ymat_fitted_raw;
SigmaE_raw_fdy = cov(resid_raw_fdy');

%% --------  Recompute the analysis to get confidence limits -------- 
% ------- Scalar responses
stderrStruct_vecy = fRegress_stderr(fRegressStruct_vecy, eye(n_train), SigmaE_vecy);
betastderrcell_vecy = stderrStruct_vecy.betastderr;

stderrStruct_raw_vecy = ...
    fRegress_stderr(fRegressStruct_raw_vecy, eye(n_train), SigmaE_raw_vecy);
betastderrcell_raw_vecy = stderrStruct_raw_vecy.betastderr;

% Constant  coefficient standard error:
intercept_std_vecy = getcoef(betastderrcell_vecy{1});
intercept_ste_raw_vecy = getcoef(betastderrcell_raw_vecy{1});

% -------- Functional responses
stderrStruct_fdy = fRegress_stderr(fRegressStruct_fdy, y2cMap, SigmaE_fdy);
% fixed a bug in fRegress_stderr.m at line 124: 
% bstderrfdj = data2fd(bstderrj, tfine, betabasisj); should be 
% bstderrfdj = data2fd(tfine, bstderrj, betabasisj);
betastderrcell_fdy = stderrStruct_fdy.betastderr;

stderrStruct_raw_fdy = ...
    fRegress_stderr(fRegressStruct_raw_fdy, y2cMap, SigmaE_raw_fdy);
betastderrcell_raw_fdy = stderrStruct_raw_fdy.betastderr;

%  Coefficient standard error:
intercept_std_fdy = eval_fd(pgrid, betastderrcell_fdy{1});
intercept_std_raw_fdy = eval_fd(pgrid, betastderrcell_raw_fdy{1});

%% -------- Predict on test data -------- 
%Set up xfd for test data
xfd_test = smooth_basis(pgrid, Xtest, xbasis);
xfd_raw_test = smooth_basis(pgrid, Xraw_test, xbasis);

% -------- Scalar responses
xfdcell_test = cell(1, 2);
xfdcell_test{1} = fd(ones(1, n_test), conbasis);
xfdcell_test{2} = xfd_test;

xfd_raw_test_cell = cell(1, 2);
xfd_raw_test_cell{1} = fd(ones(1, n_test), conbasis);
xfd_raw_test_cell{2} = xfd_raw_test;

Avec_pred = fRegressPred(xfdcell_test, betaestcell_vecy);
Avec_pred_raw = fRegressPred(xfd_raw_test_cell, betaestcell_raw_vecy);

display(['Scalar responses predict mse = ', ...
    num2str(mse(Avec_test_true, Avec_pred)), ...
    '; Scalar responses with raw data predict mse_raw = ',...
    num2str(mse(Avec_test_true, Avec_pred_raw))])

% -------- Functional responses
ymat_test_pred = ...
    eval_fd(pgrid, fRegressPred(xfdcell_test, betaestcell_fdy, xbasis));
ymat_test_pred_raw = ...
    eval_fd(pgrid, fRegressPred(xfd_raw_test_cell, betaestcell_raw_fdy, xbasis));

display(['Functional response prediction mse = ', ...
    num2str(mse(ymat_test_true, ymat_test_pred)), ...
    '; Functional responses prediction with Raw data mse_raw = ',...
    num2str(mse(ymat_test_true, ymat_test_pred_raw))])
\end{verbatim}

\end{appendices}